\documentclass[aps,floatfix,showpacs,twocolumn,superscriptaddress]{revtex4} 
 
\usepackage{graphicx} 
\usepackage{dcolumn} 
\usepackage{amsmath}

\newcommand{\sfrac}[2]{\mbox{\footnotesize $\displaystyle \frac{#1}{#2}$}}

\begin{document} 
 
\preprint{\parbox[t]{45mm}{\small ANL-PHY-10829-TH-2004\\%
                                  KSUCNR-204-01\\%
                                  MPG-VT-UR 245/04}} 
 
\title{%
\hspace*{\fill}{\tt\normalsize ANL-PHY-10829-TH-2004, KSUCNR-204-01, MPG-VT-UR 245/04}\\[1ex] 
%
Aspects and consequences of a dressed-quark-gluon vertex}

\author{M.S.\ Bhagwat} 
\affiliation{Center for Nuclear Research, Department of Physics, 
             Kent State University, Kent, Ohio 44242 U.S.A.} 

\author{A.\ H\"oll} 
\affiliation{Physics Division, Argonne National Laboratory, 
             Argonne, IL 60439-4843 U.S.A.} 
             
\author{A.\ Krassnigg}
\affiliation{Physics Division, Argonne National Laboratory, 
             Argonne, IL 60439-4843 U.S.A.} 
                          
\author{C.D.\ Roberts} 
\affiliation{Physics Division, Argonne National Laboratory, 
             Argonne, IL 60439-4843 U.S.A.} 
\affiliation{Fachbereich Physik, Universit\"at Rostock, D-18051 Rostock, 
Germany} 

\author{P.C.\ Tandy} 
\affiliation{Center for Nuclear Research, Department of Physics, 
             Kent State University, Kent, Ohio 44242 U.S.A.} 
              
\begin{abstract} 
\rule{0ex}{3ex} 
Features of the dressed-quark-gluon vertex and their role in the gap and Bethe-Salpeter equations are explored.  It is argued that quenched lattice data indicate the existence of net attraction in the colour-octet projection of the quark-antiquark scattering kernel.  This attraction affects the uniformity with which solutions of truncated equations converge pointwise to solutions of the complete gap and vertex equations.  For current-quark masses less than the scale set by dynamical chiral symmetry breaking, the dependence of the dressed-quark-gluon vertex on the current-quark mass is weak.  The study employs a vertex model whose diagrammatic content is explicitly enumerable. That enables the systematic construction of a vertex-consistent Bethe-Salpeter kernel and thereby an exploration of the consequences for the strong interaction spectrum of attraction in the colour-octet channel.  With rising current-quark mass the rainbow-ladder truncation is shown to provide an increasingly accurate estimate of a bound state's mass.  Moreover, the calculated splitting between vector and pseudoscalar meson masses vanishes as the current-quark mass increases, which argues for the mass of the pseudoscalar partner of the $\Upsilon(1S)$ to be above $9.4\,$GeV.  The absence of colour-antitriplet diquarks from the strong interaction spectrum is contingent upon the net amount of attraction in the octet projected quark-antiquark scattering kernel.  There is a window within which diquarks appear.  The amount of attraction suggested by lattice results is outside this domain.
\end{abstract} 
 
\maketitle 
 
\section{Introduction} 
\label{sec:one} 
It is difficult to overestimate the importance of dynamical chiral symmetry breaking (DCSB) in determining features of the strong interaction spectrum.  For example, this phenomenon, which can even be expressed in models that omit confinement \cite{papa1,haweswilliams,hawesmaris}, is responsible for the remarkably small value of the ratio of $\pi$- and $\rho$-meson masses and, inseparable from this, the generation of large constituent-like masses for dressed-quarks.  It follows both: that a sum of constituent-like dressed-quark masses sets a baseline for the masses of all light hadrons, except the pseudoscalar mesons; and that pseudoscalar ``meson cloud'' contributions are essential to an understanding of hadron observables, such as the masses \cite{NpiN,capstickmorel} and form factors  \cite{AWT1,AWT2} of octet and decuplet baryons.  

That stated, our primary concern herein is the dressed-quark-gluon vertex, $\Gamma^a_\nu(q;p)$.  We are generally interested in its form, how that arises, and the ways it affects and is affected by strong interaction phenomena.

This vertex is a key element in the gap equation$^{\rm\footnotemark[1]}$\footnotetext[1]{We employ a Euclidean metric, with:  $\{\gamma_\mu,\gamma_\nu\} = 2\delta_{\mu\nu}$; $\gamma_\mu^\dagger = \gamma_\mu$; and $a \cdot b = \sum_{i=1}^4 a_i b_i$.}  
\begin{eqnarray} 
\nonumber 
\lefteqn{S(p)^{-1}  =  Z_2 \,(i\gamma\cdot p + m^{\rm bm})}\\ 
&&  +\, Z_1 \int^\Lambda_q  g^2 D_{\mu\nu}(p-q) \frac{\lambda^a}{2}\gamma_\mu 
S(q) \Gamma^a_\nu(q,p) \,, \label{gendse} 
\end{eqnarray} 
and this brings an immediate link with DCSB.  The gap equation has long been used as a tool for developing insight into the origin of DCSB, and searching for a connection between this phenomenon and confinement \cite{cdragw}.  Moreover, the vertex is essential to a valid description of bound states and therefore to realising and understanding Goldstone's theorem, in particular, and current conservation in general.  These are additional qualities whose elucidation will form a large part of our discussion. 

The gap equation consists of other elements: $D_{\mu\nu}(k)$, the renormalised dressed-gluon propagator; $m^{\rm bm}$, the $\Lambda$-dependent current-quark bare mass that appears in the Lagrangian; and $\int^\Lambda_q := \int^\Lambda d^4 q/(2\pi)^4$, which represents a translationally-invariant regularisation of the integral, with $\Lambda$ the regularisation mass-scale.  The quark-gluon-vertex and quark wave function renormalisation constants, $Z_1(\zeta^2,\Lambda^2)$ and $Z_2(\zeta^2,\Lambda^2)$ respectively, depend on the renormalisation point, the regularisation mass-scale and the gauge parameter. 

The gap equation's solution is the dressed-quark propagator.  It takes the form
\begin{eqnarray} 
\nonumber 
 S(p)^{-1} & = & i \gamma\cdot p \, A(p^2,\zeta^2) + B(p^2,\zeta^2) \\ 
& =& \frac{1}{Z(p^2,\zeta^2)}\left[ i\gamma\cdot p + M(p^2,\zeta^2)\right] \label{sinvp} 
\end{eqnarray} 
and provides direct access to the gauge invariant vacuum quark condensate 
\begin{equation} 
\label{qbq0} \,-\,\langle \bar q q \rangle_\zeta^0 = \lim_{\Lambda\to \infty} 
Z_4(\zeta^2,\Lambda^2)\, N_c \, {\rm tr}_{\rm D}\int^\Lambda_q\!
S^{0}(q,\zeta)\,,  
\end{equation} 
where $Z_4$ is the renormalisation constant for the scalar part of the quark self-energy, through which the current-quark bare-mass is related to the running mass: 
\begin{equation}
Z_2(\zeta^2,\Lambda^2) \, m^{\rm bm}(\Lambda) = Z_4(\zeta^2,\Lambda^2) \, m(\zeta)\,,
\end{equation}
${\rm tr}_D$ identifies a trace over Dirac indices alone and the superscript ``$0$'' indicates the quantity was calculated in the chiral limit, which is unambiguously defined in an asymptotically free theory:
\begin{equation} 
\label{chirallimit} Z_2(\zeta^2,\Lambda^2) \,  m^{\rm bm}(\Lambda) \equiv 
0\,, \;\; \Lambda \gg \zeta \,.
\end{equation} 
The vacuum quark condensate is a primary order parameter for DCSB \cite{langfeld}.  

It is a longstanding prediction of Dyson-Schwinger equation (DSE) studies \cite{cdragw,bastirev,reinhardrev} that the two-point Schwinger functions which characterise the propagation of QCD's elementary excitations are strongly dressed at infrared length-scales, namely, $k^2\lesssim 2\,$GeV$^2$; and it has become apparent that this feature is materially important in explaining a wide range of hadron properties \cite{pieterrev}.  

Such dressing is also a feature of the dressed-quark-gluon vertex, a three-point function, and it is certain that the infrared structure of this vertex has a big impact on properties of the gap equation's solution, such as: multiplicative renormalisability \cite{cp90,bp95,bloch03}; gauge covariance \cite{br91,zmr94,bp95}, and the existence and realisation of confinement and DCSB \cite{papa1,haweswilliams,hawesmaris,aj90,wkr91,hw91}.  For example, related vertex \textit{Ans\"atze}, which agree in the ultraviolet, can yield solutions for the dressed-quark propagator via the gap equation with completely different analytic properties and incompatible conclusions on DCSB \cite{munczek86,brw92}. 

Dyson-Schwinger equation predictions for the behaviour of dressed-gluon \cite{hauckgluon,blochgluon,kondogluon} and dressed-quark propagators \cite{mr97} have been confirmed in recent numerical simulations of lattice-regularised QCD \cite{latticegluon,latticequark}.  Indeed, detailed study provides an understanding of the circumstances in which pointwise agreement is obtained \cite{mandar,alkofermaris}.  This level of sophistication does not prevail with the dressed-quark-gluon vertex, however.  Acquiring that is a realisable contemporary goal, and it is to aspects of this task that we address ourselves herein.

We have organised our presentation as follows.  In Sec.\,\ref{sectwo}, after outlining some general properties of the dressed-quark gluon vertex, we recapitulate on a nonperturbative DSE truncation scheme \cite{truncscheme,detmold} that has already enabled some systematic study of the connection between the dressed-quark-gluon vertex and the expression of symmetries in strong interaction observables.  In doing this we are led to propose an extension of earlier work, one which facilitates an exploration of the impact that aspects of the three-gluon vertex have on hadron phenomena.
To amplify the illustrative efficacy of our analysis we introduce a simple model to describe the propagation of dressed-gluons \cite{mn83} that reduces the relevant DSEs to a set of coupled algebraic equations which, notwithstanding their simplicity, exhibit characteristics essential to the strong interaction.

In Sec.\,\ref{secthree} we capitalise on the simplicity of our model and chronicle a range of qualitative features of the dressed-quark-gluon vertex and dressed-quark propagator that are common to our model and QCD.  Of particular interest are the effects of net attraction in the colour-octet quark-antiquark scattering kernel which we are able to identify.  We follow that in Sec.\,\ref{secfour} with an analysis of the Bethe-Salpeter equations which can be constructed, consistent with the fully dressed-quark-gluon vertex, so that the Ward-Takahashi identities associated with strong interaction observables are automatically satisfied.  This property is crucial to understanding hadron properties and interactions \cite{mrt98,mishasvy,bicudo,mariscotanch,krassnigg}.  In addition, we describe the evolution of pseudoscalar and vector meson masses with growing current-quark mass.  One outcome of that is a quantitative assessment of the accuracy for meson masses of the widely used rainbow-ladder truncation, which we determine by a comparison with the masses obtained with all terms in the vertex and kernel retained.  Section \ref{secfour} also contains an analysis of colour-antitriplet diquark correlations.  Here, in particular, the role played by attraction in the colour-octet projection of the quark-antiquark scattering matrix is noteworthy.

We close our presentation with a summary in Sec.\,\ref{conclusion}.

\section{Dressed Quark-Gluon Vertex}
\label{sectwo}
\subsection{General features}  
This three-point Schwinger function can be calculated in perturbation theory but, since we are interested in the role it plays in connection with confinement, DCSB and bound state properties, that is inadequate for our purposes: these phenomena are essentially nonperturbative.  

Instead we begin by observing that the dressed vertex can be written
\begin{equation}
i \Gamma_\mu^a(p,q) = \frac{i}{2} \lambda^a \, \Gamma_\mu(p,q) =: l^a\, \Gamma_\mu(p,q)\,;
\end{equation}
viz., the colour structure factorises, and that twelve Lorentz invariant functions are required to completely specify the remaining Dirac-matrix-valued function; i.e., 
\begin{eqnarray}
\nonumber \Gamma_\mu(p,q) &= & \gamma_\mu\, \hat\Gamma_1(p,q) +  \gamma\cdot (p+q)\,(p+q)_\mu \, \hat \Gamma_2(p,q) \\
&& - i (p+q)_\mu\,  \hat\Gamma_3(p,q) + [\ldots], 
\label{genvtx}
\end{eqnarray}
where the ellipsis denotes contributions from additional Dirac matrix structures that play no part herein.  Since QCD is renormalisable, the bare amplitude associated with $\gamma_\mu$ is the only function in Eq.\,(\ref{genvtx}) that exhibits an ultraviolet divergence at one-loop in perturbation theory.

The requirement that QCD's action be invariant under local colour gauge transformations entails$^{\rm\footnotemark[2]}$\footnotetext[2]{NB. Equation (\ref{STI}) is equally valid when expressed consistently in terms of bare Schwinger functions.} \cite{mp78}
\begin{eqnarray}
\nonumber
\lefteqn{k_\mu\,i\Gamma_\mu(p,q) = }\\
\nonumber && {\cal F}^g(k^2) \, \left\{ \left[ 1 - B(p,q) \right] S(p)^{-1} - S(q)^{-1} \left[ 1 - B(p,q)\right] \right\},\\
\label{STI}
\end{eqnarray}
wherein ${\cal F}^g(k^2)$, $k=p-q$, is the dressing function that appears in the renormalised covariant-gauge ghost propagator:
\begin{equation}
D^g(k^2,\zeta^2) = - \, \frac{{\cal F}^g(k^2,\zeta^2)}{k^2}\,,
\end{equation}
and $B(p,q)$ is the renormalised ghost-quark scattering kernel. At one-loop order on the domain in which perturbation theory is a valid tool 
\begin{equation}
{\cal F}^g(k^2,\zeta^2) = \left[ \frac{\alpha(k^2)}{\alpha(\zeta^2)} \right]^{\gamma_g/\beta_1},
\end{equation}
with the anomalous dimensions $\gamma_g= - (3/8) \,C_2(G)$ in Landau gauge, which we use throughout, and $\beta_1 = -(11/6)\, C_2(G) + (1/3) \,N_f$, where $C_2(G)=N_c$ and $N_f$ is the number of active quark flavours.  This result may be summarised as ${\cal F}^g(k^2)\approx 1$ on the perturbative domain, up to $\ln p^2/\Lambda_{\rm QCD}^2$-corrections.  In a similar sense, $B(p,q) \approx 0$ in Landau gauge on this domain.$^{\rm\footnotemark[3]}$\footnotetext[3]{An even closer analogy is a kindred result for $Z(p^2)$ in Eq.\,(\ref{sinvp}); viz., in Landau gauge $[1 - Z(p^2,\zeta^2)] \equiv 0$ at one loop in perturbation theory and hence, on the perturbative domain, corrections to this result are modulated by $\ln\ln p^2/\Lambda_{\rm QCD}^2$.  This very slow evolution is exhibited, e.g., in the numerical results of Ref.\ \cite{wkr91}.}

Equation (\ref{STI}) is a Slavnov-Taylor identity, one of a countable infinity of such relations in QCD, and it is plainly an extension of the Ward-Takahashi identity that applies to the fermion-photon vertex.  The Ward-Takahashi identity entails that in the vertex describing the coupling of a photon to a dressed-quark:
\begin{eqnarray}
\label{AQED}
\hat \Gamma_1^\gamma(p,p) &=& A(p^2,\zeta^2) \,,\\
\hat \Gamma_2^\gamma(p,p) &= &2\, \frac{d}{dp^2}\, A(p^2,\zeta^2) \,,\\
\label{a3QED}
\hat \Gamma_3^\gamma(p,p) &= &2\, \frac{d}{dp^2}\, B(p^2,\zeta^2) \,.
\end{eqnarray}
Identifying and understanding this nontrivial structure of the dressed-quark-photon vertex has been crucial in describing the electromagnetic properties of mesons \cite{cdrpion,pmpion,pmkaon,pmtransition}.

The similarity between the Slavnov-Taylor and Ward-Takahashi identities has  immediate, important consequences.  For example, if the result  
\begin{equation}
\label{FBconstraint}
 0 < {\cal F}^g(k^2) \left[ 1 - B(p,q) \right] < \infty
\end{equation}
also prevails on the nonperturbative domain then, because of the known behaviour of the dressed-quark propagator, Eq.\,(\ref{FBconstraint}) is sufficient grounds for Eq.\,(\ref{STI}) to forecast that in the renormalised dressed-quark-gluon vertex
\begin{equation}
1 < \hat\Gamma_1(p,p) < \infty\,
\end{equation}
at infrared spacelike momenta.  This result, an echo of Eq.\,(\ref{AQED}), signals that the complete kernel in the DSE satisfied by $\Gamma_\mu^a(p,q)$ is attractive on the nonperturbative domain.  

\subsection{Vertex in the gap equation}
The ability to use the gap equation to make robust statements about DCSB rests on the existence of a systematic nonperturbative and chiral symmetry preserving truncation scheme.  One such scheme was introduced in Ref.\ \cite{truncscheme}.  It may be described as a dressed-loop expansion of the dressed-quark-gluon vertex wherever it appears in the half amputated dressed-quark-antiquark (or -quark-quark) scattering matrix: $S^2 K$, a renormalisation-group invariant, where $K$ is the dressed-quark-antiquark (or -quark-quark) scattering kernel.  Thereafter, all $n$-point functions involved in connecting two particular quark-gluon vertices are \textit{fully dressed}. 

The effect of this truncation in the gap equation, Eq.\,(\ref{gendse}), is realised through the following representation of the dressed-quark-gluon vertex \cite{detmold}
\begin{equation}
\label{vtxexpand}
Z_1 \, \Gamma_\mu(k,p)  =  \gamma_\mu + {\cal L}_2^-(k,p) + {\cal L}_2^+(k,p) + [\ldots]\,,
\end{equation}
with
\begin{eqnarray} 
\nonumber 
\lefteqn{ {\cal L}_2^-(k,p) = \sfrac{1}{2 N_c} \int_\ell^\Lambda\! g^2 D_{\rho\sigma}(p-\ell) }\\
&& \times
\gamma_\rho S(\ell+k-p) \gamma_\mu S(\ell) 
\gamma_\sigma \,, \label{L2m}\\ 
\nonumber 
\lefteqn{ {\cal L}_2^+(k,p) =  \sfrac{N_c}{2}\int_\ell^\Lambda\! g^2\, 
D_{\sigma^\prime \sigma}(\ell) \, D_{\tau^\prime\tau}(\ell+k-p)\, }\\
 & & \times \, \gamma_{\tau^\prime} \, S(p-\ell)\, 
\gamma_{\sigma^\prime}\, 
\Gamma^{3g}_{\sigma\tau\mu}(\ell,-k,k-p) \label{L2p}\,,
\end{eqnarray}  
wherein $\Gamma^{3g}$ is the dressed-three-gluon vertex.  It is apparent that the lowest order contribution to each term written explicitly is O$(g^2)$.  The ellipsis in Eq.\,(\ref{vtxexpand}) represents terms whose leading contribution is O$(g^4)$; e.g., crossed-box and two-rung dressed-gluon ladder diagrams, and also terms of higher leading-order.

The ${\cal L}_2^-$ term in Eq.\,(\ref{vtxexpand}) only differs from a kindred term in QED by the colour factor.  However, that factor is significant because it flips the sign of the interaction in this channel with respect to QED; i.e., since 
\begin{eqnarray}
\nonumber
&&  l^a \,l^b \,l^a = \left\{\frac{1}{2} \, C_2(G)- \, C_2(R) \right\} l^b = \frac{1}{2 N_c}\, l^b \\
{\rm cf.} && l^a \, \mbox{\boldmath $1$}_c\, l^a = -\, C_2(R) \, \mbox{\boldmath $1$}_c= - \frac{N_c^2-1}{2 N_c} \, \mbox{\boldmath $1$}_c\,, \label{repulsive}
\end{eqnarray}
then single gluon exchange between a quark and antiquark is repulsive in the colour-octet channel.  Attraction in the octet channel is provided by the ${\cal L}_2^+$ term in Eq.\,(\ref{vtxexpand}), which involves the three-gluon vertex.  These observations emphasise that Eq.\,(\ref{STI}) cannot be satisfied if the contribution from the three gluon vertex is neglected because the Slavnov-Taylor identity signals unambiguously that on the perturbative domain there is net attraction in the octet channel.  

It is apparent, too, that the term involving the three gluon vertex is numerically amplified by a factor of $N_c^2\,$ cf.\ the ${\cal L}_2^-$ (Abelian-like) vertex correction.  Hence, if the integrals are of similar magnitude then the $N_c^2$-enhanced three-gluon term must dominate in the octet channel.  This expectation is borne out by the one-loop perturbative calculation of the two integrals exhibited in Eqs.\,(\ref{L2m}), (\ref{L2p})  and, moreover, the sum of both terms is precisely that combination necessary to satisfy the Slavnov-Taylor identity at this order \cite{davydychev}.

\subsection{Vertex model}
In illustrating features of the nonperturbative and symmetry preserving DSE truncation scheme introduced in Ref.\,\cite{truncscheme} in connection, for example, with DCSB, confinement and bound state structure, Ref.\,\cite{detmold} employed a dressed-quark-gluon vertex obtained by resumming a subclass of diagrams based on ${\cal L}_2^-$ alone; namely, the vertex obtained as a solution of
\begin{eqnarray}
\nonumber
\Gamma_\mu^-(k_+,k_-) &=& Z_1^{-1} \gamma_\mu + \sfrac{1}{2 N_c} \int_\ell^\Lambda\! g^2 D_{\rho\sigma}(p-\ell) \\
&& \times\,
\gamma_\rho S(\ell_+) \Gamma_\mu^-(\ell_+,\ell_-)S(\ell_-) 
\gamma_\sigma \,. \label{G2m} 
\end{eqnarray}
It was acknowledged that this subclass of diagrams is $1/N_c$-suppressed but, in the absence of  nonperturbative information about ${\cal L}_2^+$ in general, and the dressed-three-gluon vertex in particular, this limitation was accepted. 

Herein we explore a model that qualitatively ameliorates this defect while preserving characteristics that make calculations tractable and results transparent; viz., in Eq.\,(\ref{vtxexpand}) we write
\begin{equation}
{\cal L}_2^{-}+{\cal L}_2^{+} \approx {\cal L}_2^{\cal C}\,,\\
\end{equation}   
where 
\begin{eqnarray}
\nonumber\lefteqn{
{\cal L}_2^{\cal C}(k,p):= -\, {\cal C}\, C_2(R)\,  \int_\ell^\Lambda\! g^2 D_{\rho\sigma}(p-\ell)}\\
&& 
\mbox{\rule{7em}{0ex}} \times\,
\gamma_\rho S(\ell+k-p) \gamma_\mu S(\ell) 
\gamma_\sigma \,,  \label{L2C} 
\end{eqnarray}
and work with the vertex obtained as the solution of
\begin{eqnarray}
\nonumber \lefteqn{
\Gamma_\mu^{\cal C}(k_+,k_-) = Z_1^{-1} \gamma_\mu  -\, {\cal C}\, C_2(R)\, \int_\ell^\Lambda\! g^2 D_{\rho\sigma}(p-\ell)
}\\
& & \mbox{\rule{5em}{0ex}} \times\,
\gamma_\rho S(\ell_+) \Gamma_\mu^{\cal C}(\ell_+,\ell_-)S(\ell_-) 
\gamma_\sigma \,. \label{G2C} 
\end{eqnarray}

To explain this model we remark that the parameter ${\cal C}$ is a global coupling strength modifier.  (NB.\ The value ${\cal C}= -(1/8)$ reproduces the vertex resummed in Ref.\,\cite{detmold}.)  It is introduced so that our \textit{Ansatz} may \textit{mimic} the effects of attraction in the colour-octet channel without specifying a detailed form for the three-gluon vertex.  This expedient will give a faithful model so long as the integrals over the momentum dependence of ${\cal L}_2^-$ and ${\cal L}_2^+$ that appear in our calculations are not too dissimilar.  This is plausible because: they are both one-loop integrals projected onto the same Dirac and Lorentz structure and hence are pointwise similar at this order in perturbation theory; and their direct sum must conspire to give the simple momentum dependence in Eq.\,(\ref{STI}).  Moreover, as we shall demonstrate, the model has material illustrative capacity and that alone is sufficient justification for proceeding. 

As we have already noted, the value ${\cal C}= -(1/8)$ corresponds to completely neglecting the three-gluon vertex term.  There is also another particular reference case; namely, ${\cal C}=1$.  In this case a dressed-quark propagator obtained as the solution of the rainbow truncation of the gap equation:
\begin{eqnarray}
\nonumber
\lefteqn{S_R(p)^{-1}  =  Z_2 \,(i\gamma\cdot p + m^{\rm bm})}\\ 
&&  +\, \int^\Lambda_q  \! g^2 D_{\mu\nu}(p-q) \,\frac{\lambda^a}{2}\gamma_\mu \,S_R(q) \, \frac{\lambda^a}{2} \gamma_\nu \,, \label{ladderdse} 
\end{eqnarray} 
when inserted in Eq.\,(\ref{G2C}), yields a vertex $\Gamma_\mu^{{\cal C}R}(k,p)$ that satisfies
\begin{equation}
\label{GCR}
(k-p)_\mu \, i\Gamma_\mu^{{\cal C}R}(k,p) = S_R(k)^{-1} - S_R(p)^{-1}\,;
\end{equation}
viz., a Ward-Takahashi identity.  This is not materially useful, however, because herein we will seek and work with a dressed-quark propagator that is not a solution of Eq.\,(\ref{ladderdse}) but rather a solution of Eq.\,(\ref{gendse}) with a fully dressed vertex, and that vertex, determined self consistently, will not in general satisfy Eq.\,(\ref{GCR}).

\subsection{Interaction model}
A simplification, important to our further analysis, is a confining model of the dressed-gluon interaction in Eq.\,(\ref{G2C}).  We use \cite{mn83}
\begin{equation} 
\label{mnmodel} {\cal D}_{\mu\nu}(k):= g^2 \, D_{\mu\nu}(k) = 
\left(\delta_{\mu\nu} - \frac{k_\mu k_\nu}{k^2}\right) (2\pi)^4\, {\cal G}^2 \, 
\delta^4(k)\,. 
\end{equation} 
The constant ${\cal G}$ sets the model's mass-scale and henceforth we mainly set ${\cal G}=1$ so that all mass-dimensioned quantities are measured in units of ${\cal G}$.  Furthermore, since the model is ultraviolet-finite, we will usually remove the regularisation mass-scale to infinity and set the renormalisation constants equal to one.  

In these things we follow Ref.\,\cite{detmold}.  In addition, we reiterate that the model defined by Eq.\,(\ref{mnmodel}) is a precursor to an efficacious class of models that employ a renormalisation-group-improved effective interaction and whose contemporary application is reviewed in Refs.~\cite{bastirev,reinhardrev,pieterrev}.  It has many features in common with that class and, in addition, its distinctive momentum-dependence works to advantage in reducing integral equations to algebraic equations that preserve the character of the original.  There is naturally a drawback: the simple momentum dependence also leads to some model-dependent artefacts, but they are easily identified and hence not cause for concern.

\subsection{Algebraic vertex and gap equations}
If Eq.\,(\ref{mnmodel}) is used in Eq.\,(\ref{G2C}) then that part of the vertex which acts in the gap equation has no dependence on the total momentum of the quark-antiquark pair; i.e., only $\Gamma_\mu^{\cal C}(p):=\Gamma_\mu^{\cal C}(p,p)$ contributes.  In this case just the terms written explicitly in Eq.\,(\ref{genvtx}) are supported in our model for the dressed vertex, which can be expressed
\begin{equation}
\label{GCa123}
\Gamma_\mu^{\cal C}(p) = \gamma_\mu \, \alpha_1^{\cal C}(p^2) + \gamma\cdot p \,p_\mu\, \alpha_2^{\cal C}(p^2) - i p_\mu \, \alpha_3^{\cal C}(p^2)\,.
\end{equation}
This is not a severe handicap because these Dirac structures are precisely those which dominate in Eq.\,(\ref{STI}) if $B(p,q)$ is not materially enhanced nonperturbatively.  The vertex equation is
\begin{equation}
\label{vtxalgebraic} \Gamma_\mu^{\cal C}(p) = \gamma_\mu - {\cal C}\,\gamma_\rho\, 
S(p)\, \Gamma_\mu^{\cal C}(p)\, S(p)\, \gamma_\rho\,, 
\end{equation}
where we have used the fact that $C_2(R)=4/3$ when $N_c=3$.

Our analysis now mirrors that of Ref.\,\cite{detmold}.  The solution of Eq.\,(\ref{vtxalgebraic}) is:
\begin{eqnarray} 
\label{vtxsummeda} 
\lefteqn{\Gamma_\mu^{\cal C}(p) = \sum_{i=0}^\infty\,\Gamma_{\mu,i}^{\cal C}(p)} \\ 
\nonumber &=&  \sum_{i=0}^\infty\, \left[ \gamma_\mu \, \alpha^{\cal C}_{1 , i}(p^2)\,  + \, \gamma\cdot p\,p_\mu\, \alpha^{\cal C}_{ 2, i}(p^2) - \,  i 
\,p_\mu \, \alpha^{\cal C}_{3,i}(p^2)\, \right],\\ 
&& \label{vtxi} 
\end{eqnarray}  
where $\Gamma_{\mu, i}^{\cal C}(p)$ satisfies a recursion relation:
\begin{equation} 
\label{vtxrecurs} \Gamma_{\rho, i+1}^{\cal C}(p) = -{\cal C}\,\gamma_\mu\, S(p)\,  \Gamma_{\rho, i}^{\cal C}(p) \, S(p)\, \gamma_\mu\,,
\end{equation} 
with $\Gamma_{\mu, i=0 }^{\cal C} = \gamma_\mu$, the bare vertex, so that
\begin{equation}
\alpha_{1,0}^{\cal C} = 1\,,\; 
\alpha_{2,0}^{\cal C} = 0 = \alpha_{3,0}^{\cal C}\,. 
\end{equation}

In concert with Eq.\,(\ref{vtxi}), Eq.\,(\ref{vtxrecurs}) yields an algebraic matrix equation $(s=p^2)$
\begin{equation}
\mbox{\boldmath $\alpha$}^{\cal C}_{i+1}(s):= 
\left( 
\begin{array}{l} 
\rule{0ex}{2.5ex}\alpha^{\cal C}_{1,i+1}(s) \\\rule{0ex}{2.5ex} 
\alpha^{\cal C}_{2,i+1}(s) \\ 
\rule{0ex}{2.5ex}
\alpha^{\cal C}_{3 , i+1}(s) 
\end{array}\right) 
= {\cal O}^{\cal C}(s;A,B)\, \mbox{\boldmath $\alpha$}^{\cal C}_{i}(s)\,, \label{Oalpha}
\end{equation} 
where ($\Delta(s)= s A^2(s) + B^2(s)\,$)
\begin{eqnarray} 
\nonumber \lefteqn{{\cal O}^{\cal C}(s;A,B) =} \,\\ 
\nonumber && 
 - \sfrac{2 {\cal C}}{\Delta^2} \left( 
\begin{array}{ccc} 
\rule{0ex}{2.5ex} - \Delta & 0 & 0\\ 
\rule{0ex}{2.5ex} 2  A^2 & s A^2 - B^2 & 2 A B \\ 
\rule{0ex}{2.5ex} 4 A B & 4 s A B & 2 (B^2 - s A^2) 
\end{array} \right). \\ \label{Odef}
\end{eqnarray} 

It follows from Eqs.\,(\ref{vtxi}) and (\ref{Oalpha}) that
\begin{equation}
\mbox{\boldmath $\alpha$}^{\cal C} = 
\left(\sum_{i=0}^{\infty} \, \left[{\cal O}^{\cal C}\right]^i\right) \, \mbox{\boldmath $\alpha$}^{\cal C}_0 
= 
\frac{1}{1 - {\cal O}^{\cal C}}\; \mbox{\boldmath $\alpha$}^{\cal C}_0 
\end{equation}
and hence, using Eq.\,(\ref{Odef}), 
\begin{eqnarray}
\label{alpha1A}
\alpha_1^{\cal C} & = & \frac{\Delta}{\Delta - 2 {\cal C}}\,,\\
\alpha_2^{\cal C} & = & - \frac{ 4 {\cal C} A^2 }
{\Delta^2 + 2 {\cal C}(B^2 - s A^2) - 8 {\cal C}^2}
\frac{(\Delta - 4 {\cal C}) }{(\Delta - 2 {\cal C})}\,,\\
\label{alpha3A}
\alpha_3^{\cal C} & = & - \frac{ 8 {\cal C} A B }
{\Delta^2 + 2 {\cal C}(B^2 - s A^2) - 8 {\cal C}^2}\,.
\end{eqnarray}
With these equations one has a closed form for the dressed-quark-gluon vertex.  

It is evident that the momentum dependence of the vertex is completely determined by that of the dressed-quark propagator whose behaviour, however, the vertex itself influences because it appears in the gap equation:
\begin{equation}
S(p)^{-1} = i \gamma\cdot p + m +  \gamma_\mu S(p) \Gamma^{\cal C}_\mu(p)\,.
\end{equation}
Subject to Eq.\,(\ref{GCa123}), the gap equation expresses two coupled algebraic equations:
\begin{eqnarray}
\nonumber
A(s) &=& 1+\frac{1}{s A^2 + B^2} \, [ A \,(2 \alpha_1^{\cal C} - s \alpha_2^{\cal C}) - B \, \alpha_3^{\cal C}]\,,\\ \label{Asfull}
\\
\nonumber
B(s) &=& m+\frac{1}{s A^2 + B^2}\, [ B\, (4 \alpha_1^{\cal C} + s \alpha_2^{\cal C}) - sA\, \alpha_3^{\cal C}]\,.\\\label{Bsfull}
\end{eqnarray}
The dressed-quark propagator and -quark-gluon vertex follow from the solution of these equations, which is generally obtained numerically.

In the chiral limit, which here is simply implemented by setting $m=0$, a realisation of chiral symmetry in the Wigner-Weyl mode is always possible: it corresponds to the $B\equiv 0$ solution of the gap equation.  However, since the phenomena of QCD are built on a Nambu-Goldstone realisation of chiral symmetry, we do not consider the Wigner-Weyl mode any further.  Its characterisation can be achieved in a straightforward manner by adapting the analysis of Ref.\,\cite{detmold} to our improved vertex model.

\section{Propagator and vertex solutions}
\label{secthree}
\subsection{Algebraic results}
At this point some observations are useful in order to establish a context for our subsequent results.  To begin we explore the ultraviolet behaviour of the model.  It is ultraviolet finite and hence at large spacelike momenta, $s\gg 1$ (in units of ${\cal G}^2$)
\begin{eqnarray}
\label{Auv}
A(s) & \approx & 1 + \frac{\b{a}_{1}}{s} \,,\;  \\
B(s)& \approx & m +  \frac{\b{b}_{1}}{s} \,,
\label{Buv}
\end{eqnarray}
with $m$ the model's finite current-quark mass.  The model is useful because these results persist in asymptotically free theories, up to $\ln p^2/\Lambda_{\rm QCD}^2$-corrections.  With this behaviour it follows from Eqs.\,(\ref{alpha1A}) -- (\ref{Buv}) that 
\begin{eqnarray}
\label{a1mnuv}
\alpha_1^{\cal C}(s) &\approx &1 + \frac{2 {\cal C}}{s}\,,\\
\label{a2mnuv}
 \alpha_2^{\cal C}(s) &\approx& - \frac{ 4 {\cal C}}{s^2} \,,\;\\
\label{a3mnuv}
\alpha_3^{\cal C}(s) &\approx& - \, m \, \frac{ 8{\cal C} }{s^2} \,,
\end{eqnarray}
and these results in turn mean that in the ultraviolet the behaviour of the massive dressed-quark propagator is determined by the $\alpha_1$ term in the vertex, so that
\begin{equation}
\label{ABUV}
\b{a}_{1} = 2\,,\; 
\b{b}_{1} = 4\, m\,.
\end{equation}

The expansion of $\alpha_1(s)$ around $1/s=0$ reported in Eq.\,(\ref{a1mnuv}) is the same as that which arises in QCD, apart from the usual $\ln( s/\Lambda_{\rm QCD}^2)$-corrections.  However, the leading terms in $\alpha_{2,3}$ are different: on the perturbative domain in QCD these functions both begin with a term of order $(1/s)[\ln( s/\Lambda_{\rm QCD}^2)]^d$, with $d$ some combination of anomalous dimensions.  The reason for the mismatch is readily understood.  At one-loop order in QCD
\begin{equation}
B(p^2) = \hat m \, \left[\frac{1}{2} \ln (p^2/\Lambda_{\rm QCD}^2)\right]^{\gamma^m_1/\beta_1} \,,
\end{equation}
with $\hat m$ the renormalisation point invariant current-quark mass and  $\gamma^m_1 = (3/2)\, C_2(R)$.  (This makes explicit the logarithmic correction to the leading term in Eq.\,(\ref{Buv}).)  For the purpose of this explanation then, on the perturbative domain, with ${\cal F}^g(k^2) \approx 1$ and $B(p,q)\approx 0$, the Slavnov-Taylor identity, Eq.\,(\ref{STI}), is approximately equivalent to the Ward-Takahashi identity.  Hence, via Eq.\,(\ref{a3QED}),
\begin{equation}
\alpha_3(s) \approx  \frac{ \hat m }{s} \, \frac{\gamma^m_1}{\beta_1}\,
\left[ \frac{1}{2} \ln (p^2/\Lambda_{\rm QCD}^2) 
\right]^{\gamma^m_1/\beta_1-1 }\!.
\end{equation}
It is thus evident that in QCD, even though they are not themselves divergent, the leading order terms in both $\alpha_{2,3}$ are induced by the momentum-dependent renormalisation of elements contributing to their evaluation.  Such terms are naturally missing in our ultraviolet finite model.  The absence of $1/s$ terms in Eqs.\,(\ref{a2mnuv}) and (\ref{a3mnuv}) is therefore a model artefact.

We observe in addition that with attraction in the colour-octet channel; namely, for ${\cal C}>0$, $(\alpha_1^{\cal C}-1)$ is necessarily positive on the perturbative domain and $\alpha_{2,3}^{\cal C}$ are negative.  These results are also true in QCD ($\alpha_1^{\cal C}>1$ up to logarithmic corrections).  It would be an exceptional result if these statements were not also true on the nonperturbative domain.

We now turn to the infrared domain and focus on $s=0$ but consider $0<m\ll 1$, in which case 
\begin{eqnarray}
\label{Air}
A(s=0,m) & \approx & a_0^0 + a_0^{1} \,m \,,\;  \\
B(s=0,m)& \approx & b_0^0+  b_0^1 \,m  \,.
\label{Bir}
\end{eqnarray}
Upon insertion of these expressions into the gap equation one obtains
\begin{eqnarray}
\label{a00m}
&& a_0^0 = 2\, \frac{2 + 3 {\cal C}}{2 + {\cal C}}\,,\; 
a_0^1 = - \surd 2 \, \frac{4 + 20 {\cal C} + 15 {\cal C}^2}
{(2 + {\cal C})^{5/2}}\,,\\ 
&& b_0^0 = \sqrt{4 + 2\,{\cal C}}\,,\; b_0^1 = \frac{2}{(b_0^0)^2} = \frac{1}{2+{\cal C}} \,;\label{b00m}
\end{eqnarray}
viz., results which show that in the neighbourhood of $s=0$ and with attraction in the colour-octet channel, $A(s)$ decreases with increasing current-quark mass while $B(s)$ increases.  At this order the mass function is
\begin{equation}
\label{Ms0m}
M(s=0,m)= \frac{B(0,m)}{A(0,m)} =   \mu_0^0 + \mu_0^1 \, m
\end{equation}
with
\begin{eqnarray}
\mu_0^0= \frac{b_0^0}{a_0^0} &= &\frac{{\sqrt{2}}\,{\left( 2 + {\cal C} \right) }^
     {\frac{3}{2}}}{4 + 6\,{\cal C}}\,,\\
\mu_0^1 = \frac{b_0^1}{a_0^0} - \frac{b_0^0 \, a_0^1}{(a_0^0)^2}&=& \frac{6 + 23\,{\cal C} + 15\,{{\cal C}}^2}
  {2\,{\left( 2 + 3\,{\cal C} \right) }^2}\,.
\end{eqnarray}
For ${\cal C} > -(1/3)$ the mass function also increases with rising $m$.

The infrared behaviour of the dressed-quark-gluon vertex follows immediately via Eqs.\,(\ref{alpha1A}) -- (\ref{alpha3A}); using which one finds
\begin{eqnarray}
\label{a1m0}
\alpha_i^{\cal C}(s=0,m) & \approx & {\rm a}_{i,0}^{\cal C} + {\rm a}_{i,1}^{\cal C} \, m\,,\; i=1,2,3,
%
%
\end{eqnarray}
with 
\begin{eqnarray}
{\rm a}_{1,0}^{\cal C} &= &1 + \frac{\cal C}{2} \,,\; \\
{\rm a}_{1,1}^{\cal C} &=& - \frac{1}{\surd 8} \, \frac{\cal C}{\sqrt{2 + {\cal C}}}\,,\\
{\rm a}_{2,0}^{\cal C}&=&- \, \frac{{\cal C} (2 - {\cal C}) (2 + 3 {\cal C})}{(2+{\cal C})^2}\,,\\
 {\rm a}_{2,1}^{\cal C} &=& 
 \frac{2\,{\sqrt{2}}\,{\cal C}\,
    \left( 8 + 20\,{\cal C} + 2\,{{\cal C}}^2 - 
      9\,{{\cal C}}^3 \right) }{{\left( 2 + 
       {\cal C} \right) }^{ 7/2}}\,,\\
{\rm a}_{3,0}^{\cal C}&=& - \, \frac{ \surd 8 \, {\cal C}}{\sqrt{2 + {\cal C}}}\,,\\
{\rm a}_{3,0}^{\cal C}&=&
\frac{2\,{\cal C}\,\left( 5 + 6\,{\cal C} \right) }
  {{\left( 2 + {\cal C} \right) }^2}\,.\label{a3m0}
\end{eqnarray}
These algebraic formulae provide a clear indication of the effect on the dressed-quark-gluon vertex of attraction in the projection of the quark-antiquark scattering kernel onto the colour-octet channel.  Attraction causes $\hat\Gamma_1(p,p)$ to be enhanced in the infrared cf.\ the bare vertex; and it drives $\hat\Gamma_{2,3}(p,p)$ to magnified, negative values.  These results also signal that there are no surprises in the evolution into the infrared of the ultraviolet behaviour expressed in Eqs.\,(\ref{a1mnuv}) -- (\ref{a3mnuv}): in each case attraction ensures that a current-quark mass acts to reduce the vertex function in magnitude.

\subsection{Numerical results}
We have not hitherto specified a value for the parameter ${\cal C}$.  Remember, a positive value ensures net attraction in the colour-octet quark-antiquark scattering kernel and thereby models the dominance of the intrinsically non-Abelian ${\cal L}_2^+$ contribution to the dressed-quark-gluon vertex over the Abelian-like term  ${\cal L}_2^-$.  

In choosing a value for ${\cal C}$ we elect to be guided by results from  quenched lattice-QCD simulations of the dressed-quark propagator \cite{latticequark} and dressed-quark gluon vertex \cite{skullerud}.  We focus on a current-quark mass common to both simulations; namely, $60\,$MeV, at which value the lattice dressed-quark propagator has \cite{mandar}: $Z_{\rm qu}(0)\approx 0.7$, $M_{\rm qu}(0)\approx 0.42$.  Then, so as to work with dimensionless quantities, we set $m_{60} = 0.06/M_{\rm qu}(0)$ and, using Eqs.\,(\ref{Air}) -- (\ref{a3m0}), require a least-squares fit to$^{\rm\footnotemark[4]}$\footnotetext[4]{There is a confusion of positive and negative signs in Ref.\,\cite{skullerud} concerning $\lambda_2$, $\lambda_3$, as defined therein.  Our signs are correct.  With the conventions expressed in Eq.\,(\ref{GCa123}): $4 \lambda_2 = - \alpha_2$ and $2 \lambda_3 = \alpha_3$.}
\begin{eqnarray}
A(m_{\rm 60}) & = & 1.4\,, \\
\alpha_1^{\cal C}(0,m_{60}) & = & 2.1 \,, \\
-M(0,m_{\rm 60})^2 \, \alpha_2^{\cal C}(0,m_{60}) &=& 7.1 \,,\label{Ca2}\\
- M(0,m_{\rm 60}) \; \alpha_3^{\cal C}(0,m_{60}) &=& 1.0 \,.
\end{eqnarray}
This procedure yields
\begin{equation}
\label{Cbar}
{\cal C} = \bar{\cal C} = 0.51\,
\end{equation}
with an average relative error $\bar r=25$\% and standard deviation $\sigma_r=70$\%.  It is plain that the fit is not perfect but, on the other hand, the model is simple.  We note that for ${\cal C}= 0.6$:, $\bar r = 21$\%, $\sigma_r= 72$\%, while for ${\cal C}= 0.4$: $\bar r = 31$\%, $\sigma_r= 67$\%.  If one omits Eq.\,(\ref{Ca2}) from the fitting requirements then $\bar{\cal C}=0.49$ with $\bar r = 2.5\,$\% and $\sigma_r=63$\%.  It is evident that competing requirements bound the amount of attraction allowed in the kernel.  We can now illustrate the results for the dressed-quark propagator and dressed-quark-gluon vertex.  

\begin{figure}[t] 
\vspace*{2em}
 
\centerline{\includegraphics[width=0.45\textwidth]{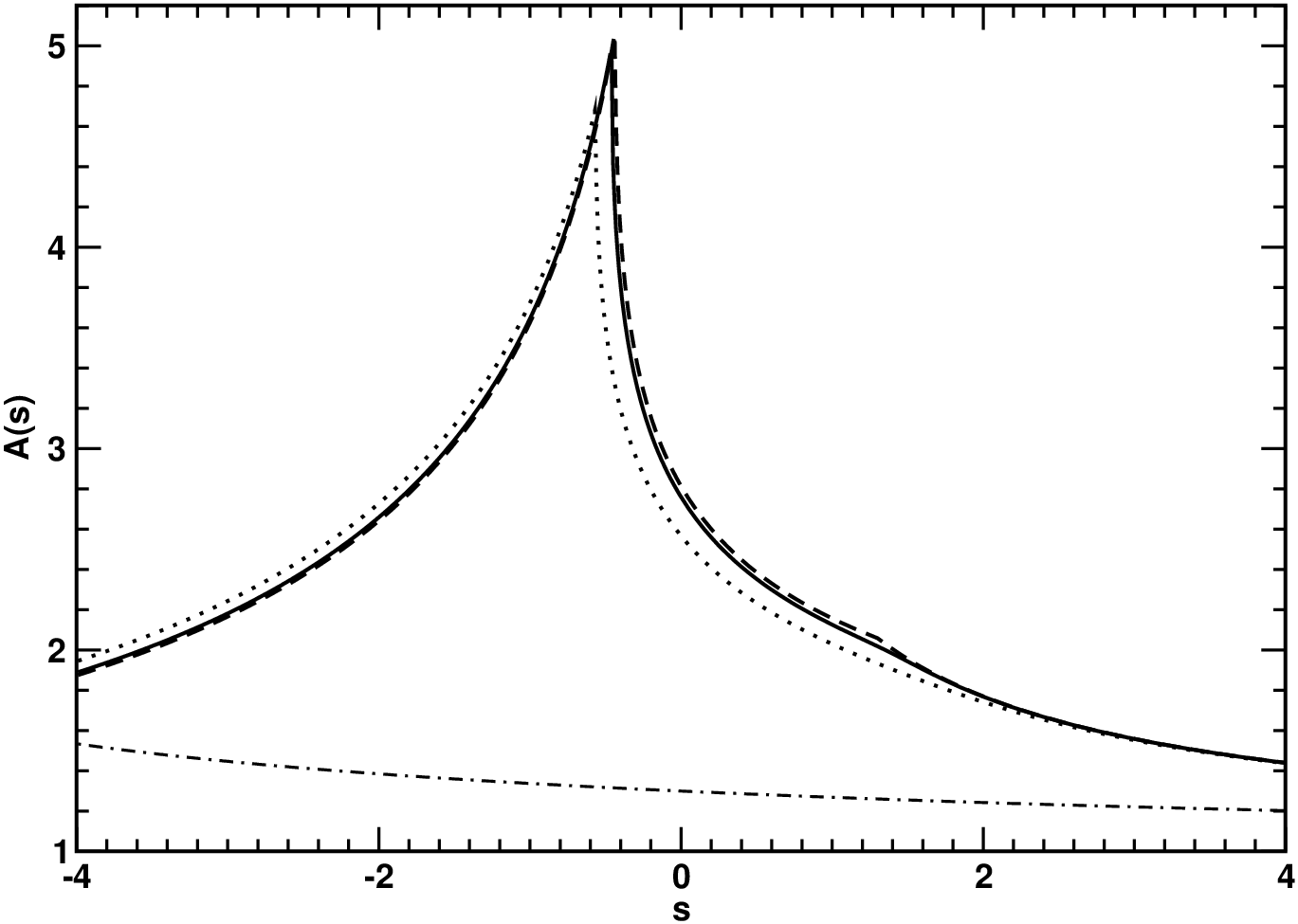}}

\caption{\label{As} Current-quark-mass-dependence of the inverse of the dressed-quark wave function renormalisation.  For all curves ${\cal C} = \bar{\cal C} = 0.51$.  Dash-dot line: $m=2$; dotted line: $m=m_{60}$; solid line: $m=0.015$; dashed line: chiral limit, $m=0$.  All dimensioned quantities are measured in units of ${\cal G}$ in Eq.\,(\protect\ref{mnmodel}).  A fit to meson observables requires ${\cal G}=0.69\,$GeV and hence $m=0.015$ corresponds to $10\,$MeV.}
\end{figure} 

\begin{figure}[t] 
\vspace*{2em}
 
\centerline{\includegraphics[width=0.46\textwidth]{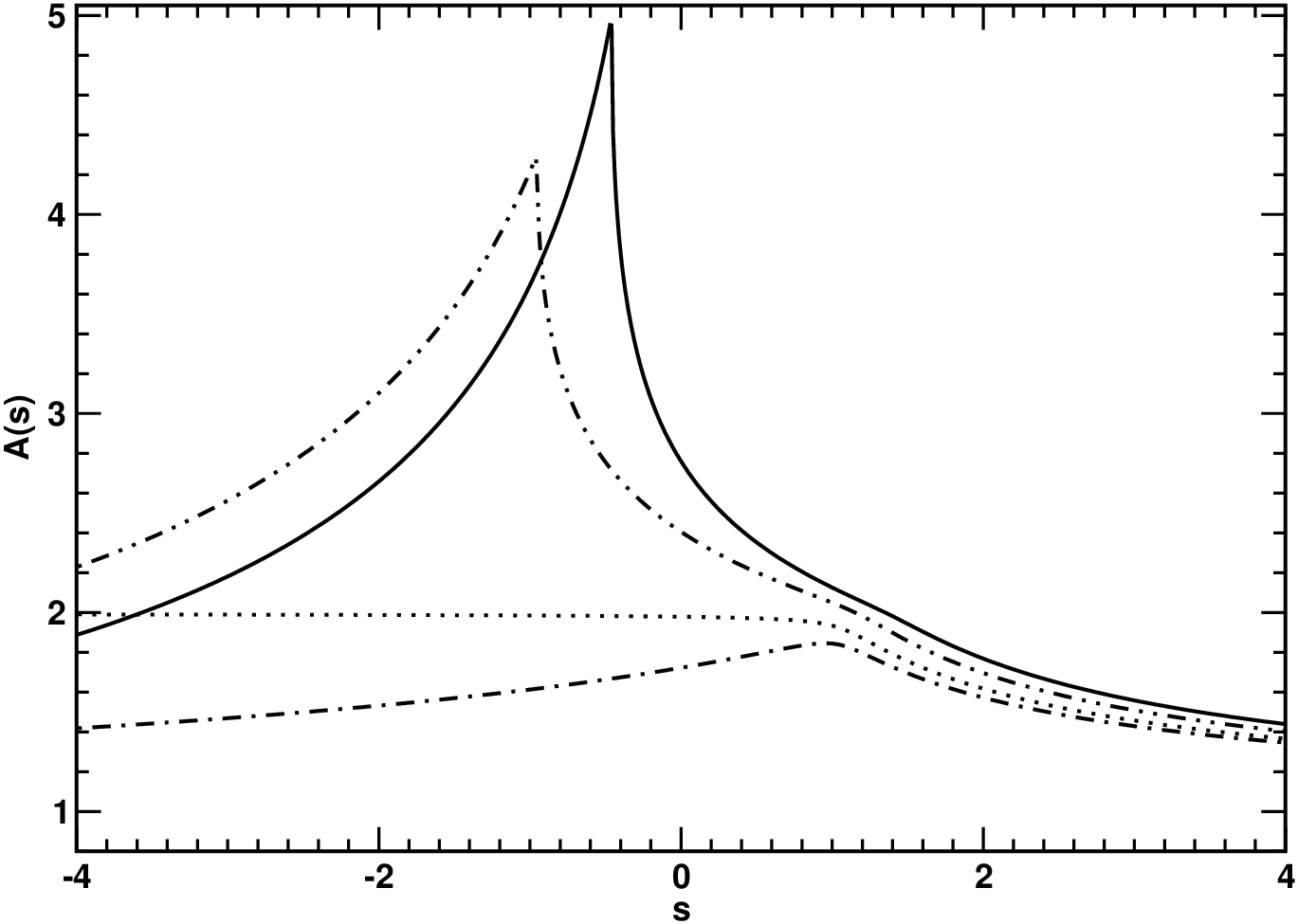}}
 
\caption{\label{As1} ${\cal C}$-dependence of $A(s)$.  For all curves $m=0.015$.  Solid line: ${\cal C}=\bar{\cal C}=0.51$; dash-dot-dot line: ${\cal C}=1/4$; dotted line: ${\cal C}=0$, which is equivalent to the rainbow truncation result first obtained in Ref.\,\protect\cite{mn83}; and dash-dot curve: ${\cal C}=-1/8$, for comparison with Ref.\,\protect\cite{detmold}.  Dimensioned quantities are measured in units of ${\cal G}$ in Eq.\,(\protect\ref{mnmodel}).}
\end{figure}

\begin{figure}[t] 
\vspace*{2em}
 
\centerline{\includegraphics[width=0.46\textwidth]{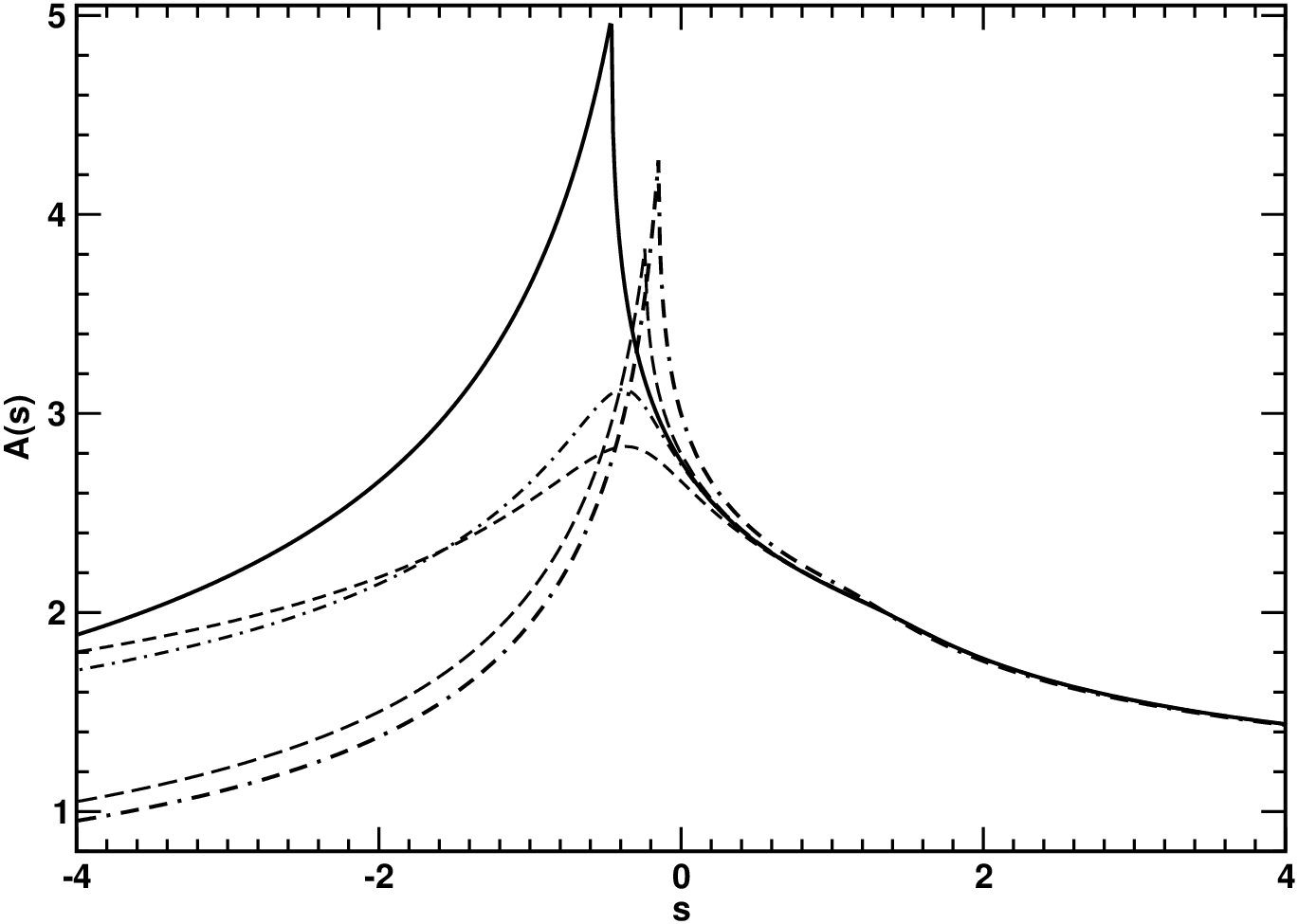}}
 
\caption{\label{Asl} Pointwise approach by solutions of the truncated gap equation to that obtained with the completely resummed vertex.  Solid line: complete solution; and in addition: dash-dash-dot line - result for $A(s)$ obtained with only the $i=0,1$ terms retained in Eq.\,(\protect\ref{vtxsummeda}), which corresponds to the one-loop corrected vertex; short-dash line - two-loop-corrected vertex, $i=0,1,2$; long-dash line - three-loop-corrected vertex; and short-dash-dot line: four-loop corrected vertex.  For all curves, ${\cal C}=\bar{\cal C}=0.51$ and $m=0.015$.  Dimensioned quantities are measured in units of ${\cal G}$ in Eq.\,(\protect\ref{mnmodel}).}
\end{figure}

In Figs.\,\ref{As} -- \ref{Asl} we depict $A(p^2)$, the scalar function which characterises the Dirac vector piece in the inverse of the dressed-quark propagator.  Figure \ref{As} shows $A(p^2)$ to evolve slowly with the current-quark mass when that mass is significantly smaller than the model's mass-scale.  However, when the current-quark mass becomes commensurate with or exceeds that mass-scale, it acts to very effectively dampen this function's momentum dependence so that $A(p^2)\approx 1$.  This is also true in QCD.  

Figure \ref{As1} exhibits the ${\cal C}$-dependence of $A(p^2)$.  A comparison of the ${\cal C}=-1/8$, $0$, $1/4$ and $0.51$ curves shows clearly that the presence of net attraction in the colour-octet quark-antiquark scattering kernel uniformly increases the magnitude of $A(p^2)$ at all momenta.  This effect is pronounced at infrared spacelike momenta and particularly on the timelike domain, $s<0$.  In this figure, as in those which follow, ${\cal C}=0$ corresponds to the rainbow-ladder DSE truncation; i.e., the leading order term of the truncation scheme introduced in Ref.\,\cite{truncscheme}.  

In Fig.\,\ref{Asl} we illustrate the effect of adding corrections to the vertex, one at a time, in the presence of the amount of attraction suggested by lattice data.  On the spacelike domain, $s>0$, the one-, two-, three- and four-loop corrected vertices yield a result for $A(p^2)$ that is little different from that produced by the completely resummed vertex (solid line).  However, that is not true on the timelike domain, whereupon confinement is expressed and hence it is unsurprising that nonperturbative effects become important.  In our model, as in the original version \cite{mn83}, confinement is realised via the absence of a particle-like singularity in the dressed-quark propagator \cite{hawesmaris}.  The cusp displayed by $A(p^2)$ in the timelike domain is one manifestation of this feature.  While it moves toward $s=0$ with increasing ${\cal C}$, it does not enter the spacelike domain.  The figure makes clear that convergence to the solution obtained with the completely resummed vertex proceeds via two routes: one followed by solutions obtained with an odd number of loop corrections to the vertex; and another by those obtained using a vertex with an even number of loop corrections.  This effect is not present with net repulsion in the colour-octet projection of the quark-antiquark scattering kernel.

\begin{figure}[t] 
\vspace*{2em}

\centerline{\includegraphics[width=0.46\textwidth]{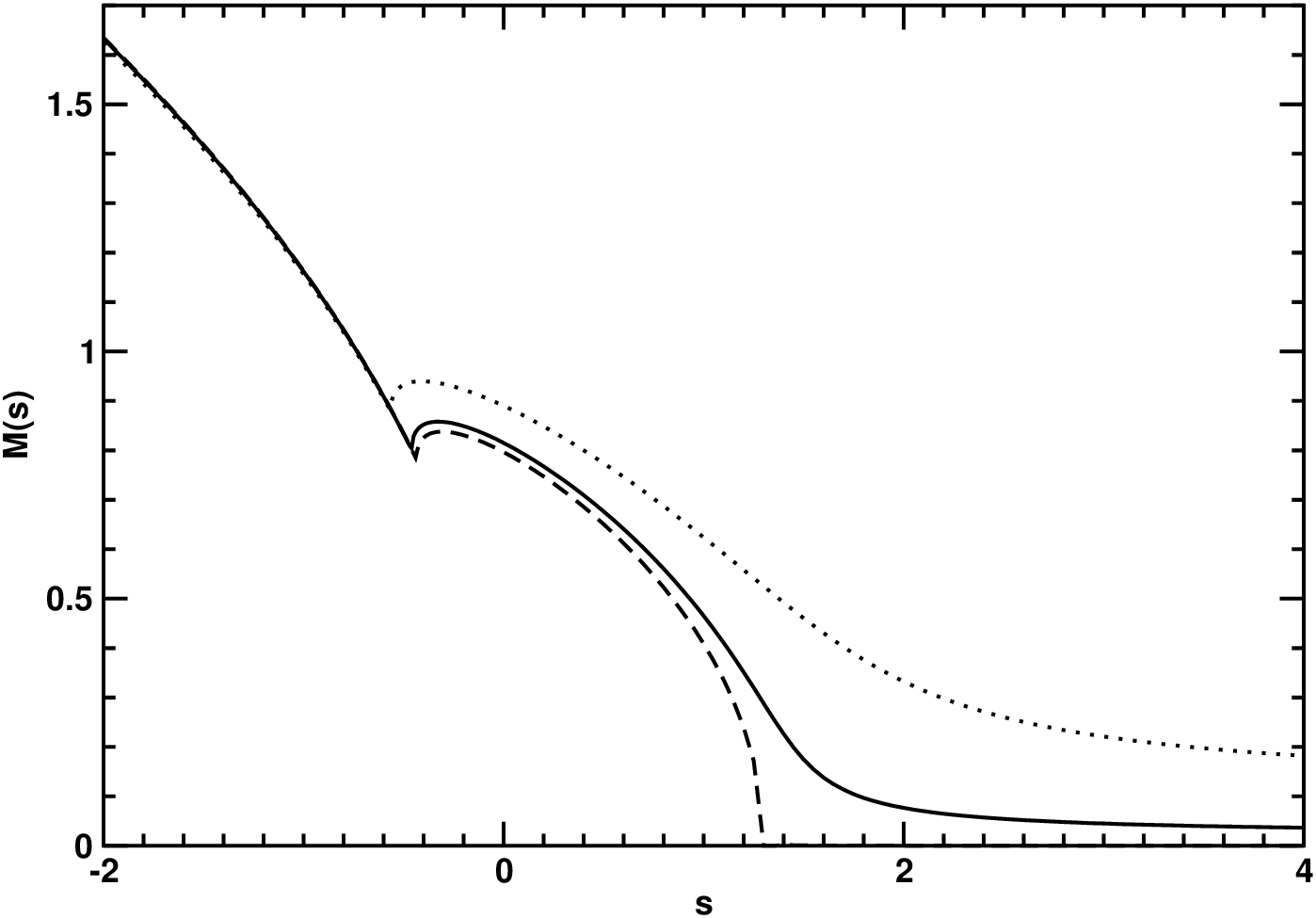}}

\vspace*{3.5em}

\centerline{\includegraphics[width=0.46\textwidth]{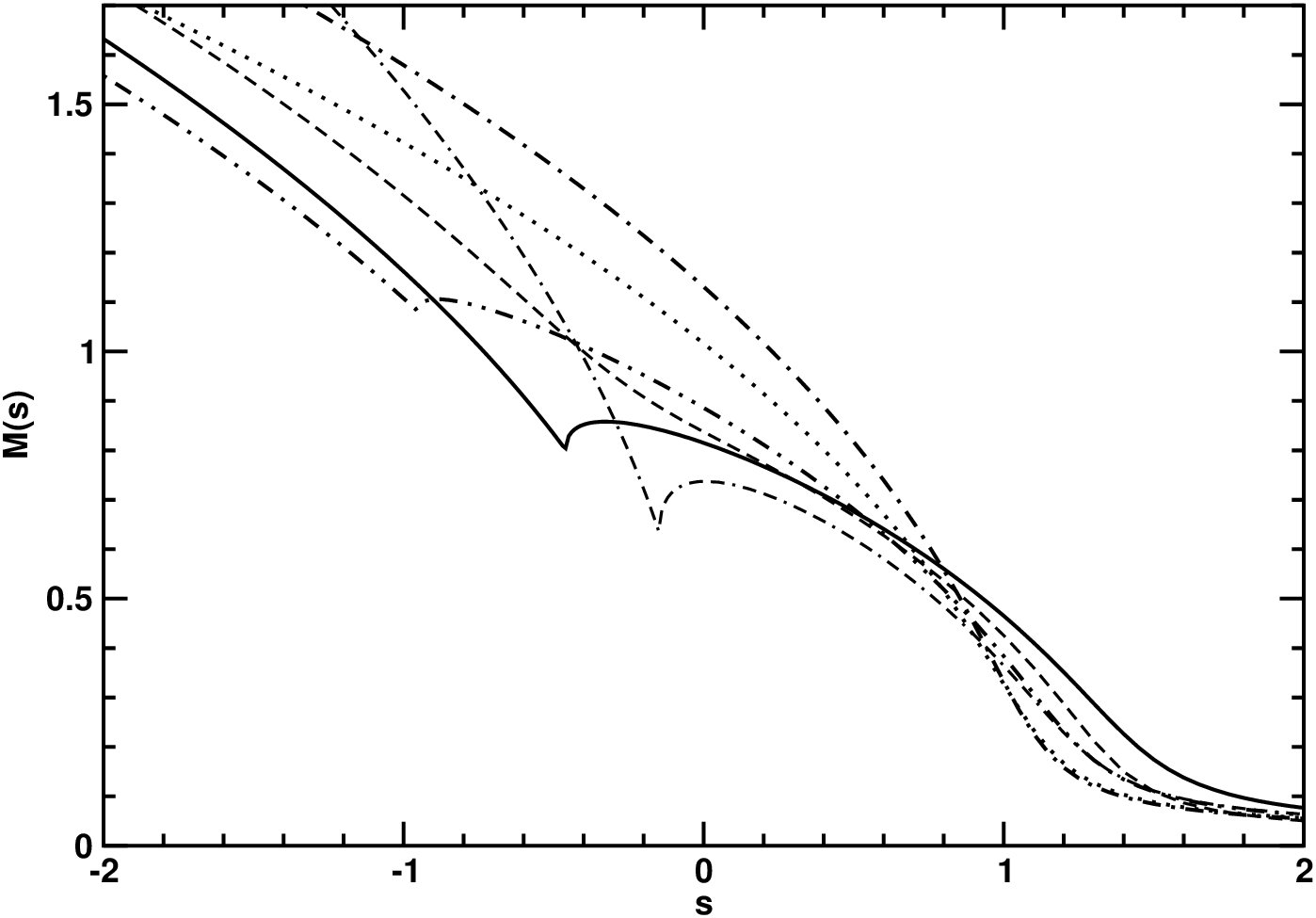}}

\caption{\label{Ms} \textit{Upper panel} -- Current-quark-mass-dependence of the dressed-quark mass function.  For all curves ${\cal C} = \bar{\cal C} = 0.51$.  Dotted line: $m=m_{60}$; solid line: $m=0.015$; dashed line: chiral limit, $m=0$.  \textit{Lower panel} -- ${\cal C}$-dependence of this function.  For all curves $m=0.015$.  Solid line: ${\cal C}=\bar{\cal C}=0.51$; dash-dot-dot line: ${\cal C}=1/4$; dotted line: ${\cal C}=0$, which is equivalent to the rainbow truncation result first obtained in Ref.\,\protect\cite{mn83}; and dash-dot curve: ${\cal C}=-1/8$, for comparison with Ref.\,\protect\cite{detmold}.  In addition, for ${\cal C}=0.51$: dash-dash-dot line - result for $M(s)$ obtained with only the $i=0,1$ terms retained in Eq.\,(\protect\ref{vtxsummeda}), which is the one-loop corrected vertex; and short-dash line - two-loop-corrected vertex.  Dimensioned quantities are measured in units of ${\cal G}$ in Eq.\,(\protect\ref{mnmodel}).}
\end{figure} 

We plot the dressed-quark mass function in Fig.\,\ref{Ms}.  The upper panel  illustrates the dependence of $M(p^2)$ on the current-quark-mass.  The existence of a nontrivial solution in the chiral limit is the realisation of DCSB, in our model and also in QCD.  For current-quark masses less than the model's mass-scale, ${\cal G}$, the dynamically generated mass determines the scale of observable quantities.  However, for $m\gtrsim {\cal G}$, this explicit chiral symmetry breaking mass-scale overwhelms the scale generated dynamically and enforces $M(p^2)\approx m$.  This is the behaviour of the $b$-quark mass function in QCD \cite{mishasvy}.  

In the lower panel of Fig.\,\ref{Ms} we depict the ${\cal C}$-dependence of $M(p^2)$.  In discussing this figure we take the rainbow-ladder result, ${\cal C}=0$, illustrated with the dotted line, as our reference point.  From a comparison of this curve with the ${\cal C}=-1/8$ result (dash-dot line) and the $\bar{\cal C}=0.51$ result (solid line) it is immediately apparent that vertex dressing driven by net attraction in the colour-octet quark-antiquark scattering kernel reduces the magnitude of the mass function at infrared momenta, a trend which is reversed for spacelike momenta $s\gtrsim {\cal G}$.$^{\rm\footnotemark[5]}$\footnotetext[5]{This pattern of behaviour is familiar from explorations \protect\cite{hw91,brw92} of the effect in the gap equation of vertex \textit{Ans\"atze} \protect\cite{cp90,bc80}; i.e., models for the vertex whose diagrammatic content is unknown but which exhibit properties in common with our calculated \mbox{${\cal C}>0$} result.}  This effect, which is correlated with the enhancement of $A(p^2)$, has an impact on the magnitude of the vacuum quark condensate.  

For example, the mapping of Eq.\,(\ref{qbq0}) into our model is:
\begin{equation}
-\langle \bar q q \rangle^0 = \frac{3}{8\pi^2}\,\int_0^{s_0}\! ds \, s \,\frac{Z(s)\,M(s)}{s+M(s)^2}\,,
\end{equation}
where $s_0$ is the spacelike point at which the model's mass function vanishes in the chiral limit, and we find
\begin{eqnarray}
\label{qbqCbar}
-\langle \bar q q \rangle^0_{{\cal C}=\bar{\cal C}} &=& (0.183 \,{\cal G})^3 =  (0.13\,{\rm GeV})^3
\end{eqnarray}
with ${\cal G}=0.69\,$GeV.  The rainbow-ladder result is $-\langle \bar q q \rangle^0_{{\cal C}=0} =  {\cal G}^3/(20 \pi^2)= (0.172 \,{\cal G})^3=(0.12\,{\rm GeV})^3$ so that
\begin{equation}
\frac{\langle \bar q q \rangle^0_{{\cal C}=0}}{\langle \bar q q \rangle^0_{{\cal C}=\bar {\cal C}}} = 0.82\,.
\end{equation}
This ratio drops to $0.50$ when ${\cal C}=1.0$ is used to calculate the denominator.  

It is thus evident that in the presence of attraction in the colour-octet quark-antiquark scattering kernel and with a common mass-scale the condensate is significantly larger than that produced by a ladder vertex
owing to an expansion of the domain upon which the dressed-quark mass function has nonzero support.

It is natural to ask for the pattern of behaviour in the presence of repulsion.  In this case Fig.\,\ref{Ms} indicates that with ${\cal C}=-1/8$ the value of the mass function is enhanced at $s=0$, a feature that is associated with the suppression of $A(s)$ which is apparent Fig.\,\ref{As1}.  The magnitude of the mass function grows larger still with a further decrease in ${\cal C}$ and its domain of nonzero support expands.  Therefore here, too, the condensate is larger than with the ladder vertex; e.g.,
\begin{equation}
\frac{\langle \bar q q \rangle^0_{{\cal C}=0}}{\langle \bar q q \rangle^0_{{\cal C}=-\frac{1}{8}}} = 0.92\,,
\end{equation}
and the ratio drops to $0.49$ when ${\cal C}= -3/8$ is used to evaluate the denominator.  

The implication of these results is that in general, with a given mass-scale and a common model dressed-gluon interaction, studies employing the rainbow-ladder truncation will materially underestimate the magnitude of DCSB relative to those that employ a well-constrained dressed-quark-gluon vertex.  Naturally, in practical phenomenology, alterations of the mass-scale can compensate for this \cite{hw91}.

Figure \ref{Ms} also shows that the pointwise convergence of $M(s)$ to the solution obtained with the fully resummed vertex, by solutions obtained with truncated vertices, follows a pattern similar to that exhibited by $A(s)$.

\begin{figure}[t] 
\vspace*{2em}
 
\centerline{\includegraphics[width=0.45\textwidth]{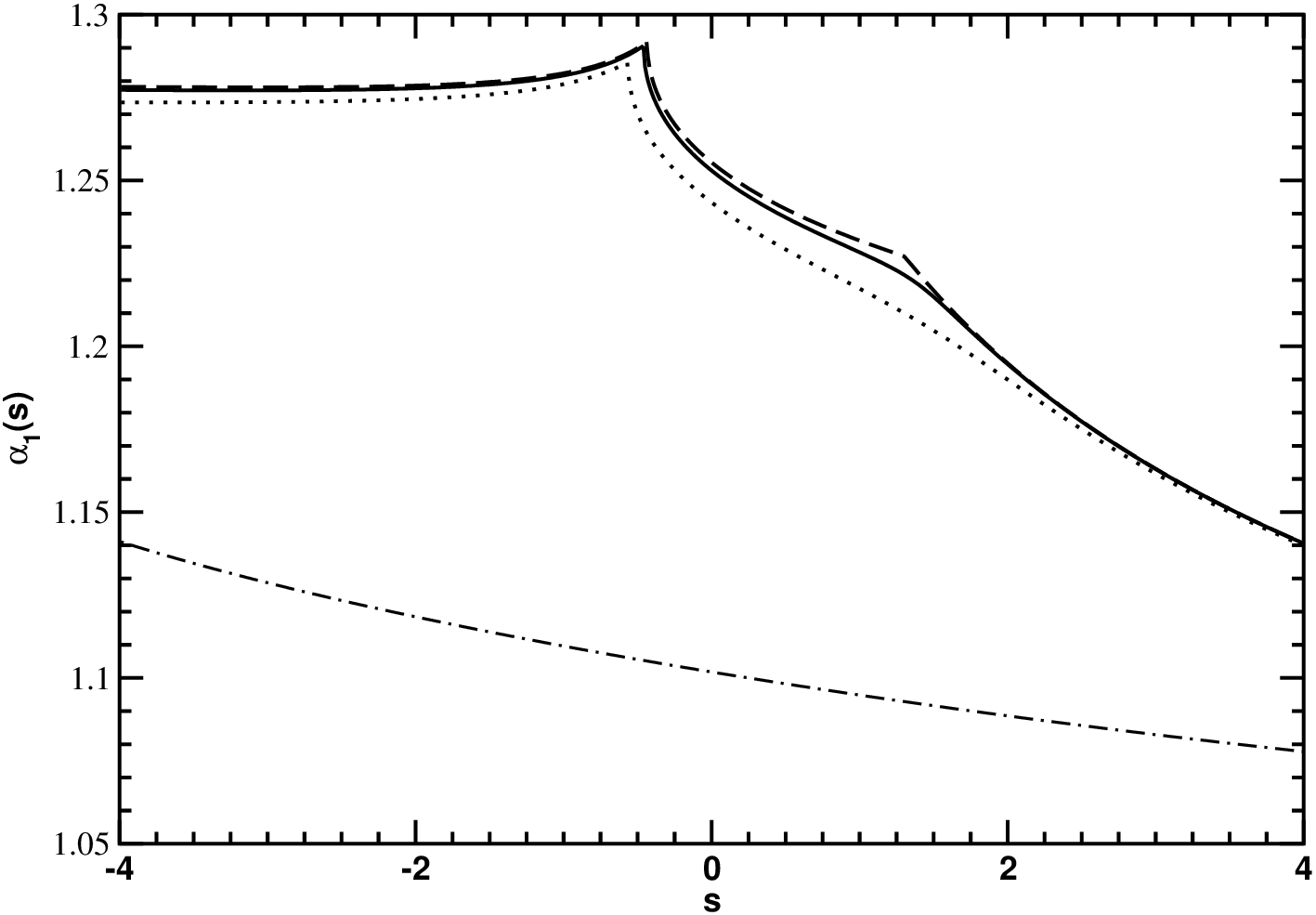}}

\vspace*{3.5em}

\centerline{\includegraphics[width=0.46\textwidth]{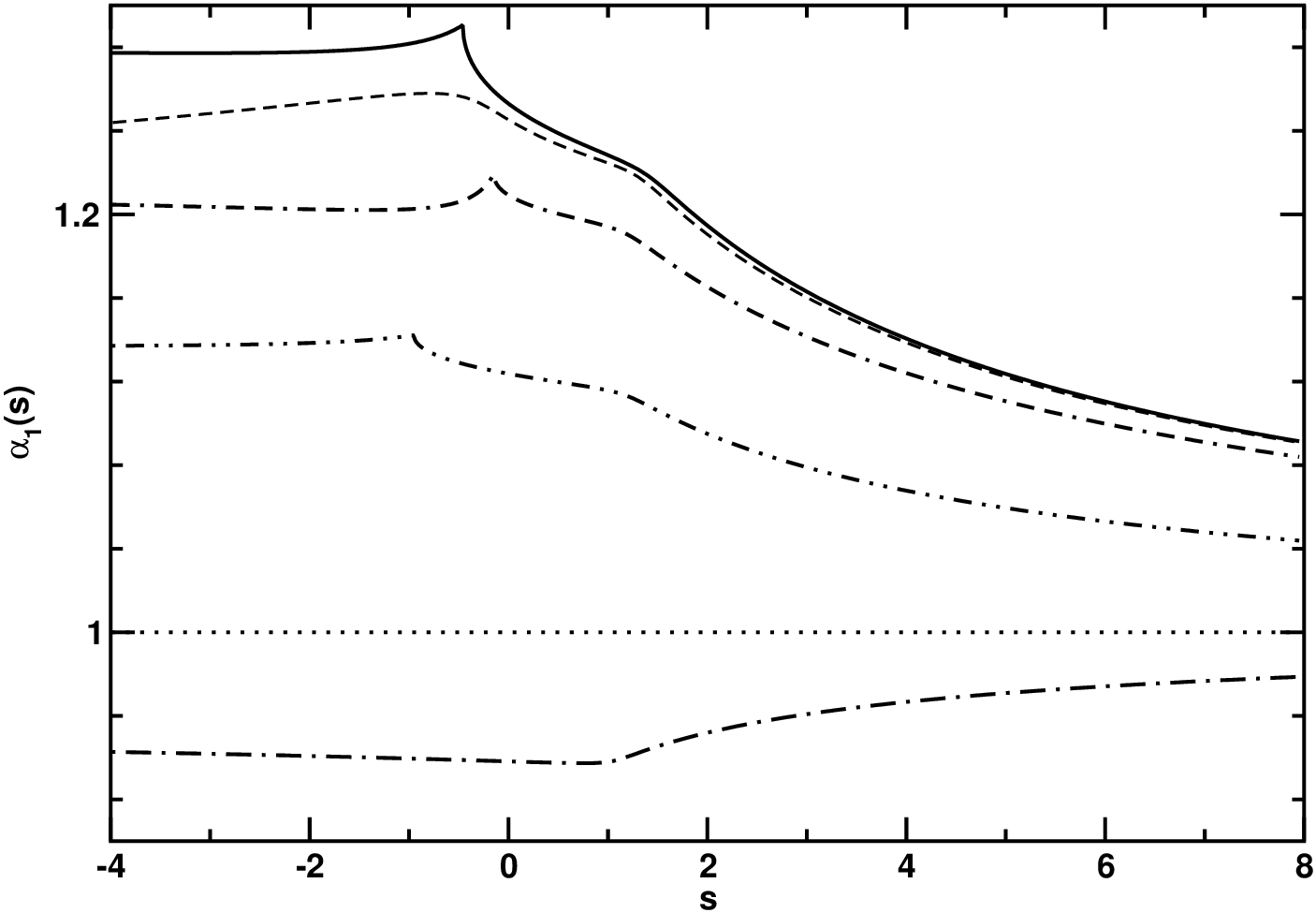}}
 
\caption{\label{a1s} \textit{Upper panel} -- Current-quark-mass-dependence of $\alpha_1^{\cal C}(s)$; viz., the leading component in the dressed-quark-gluon vertex.  For all curves ${\cal C} = \bar{\cal C} = 0.51$.  Dash-dot line: $m=2$; dotted line: $m=m_{60}$; solid line: $m=0.015$; dashed line: chiral limit, $m=0$.  \textit{Lower panel} -- ${\cal C}$-dependence of this function.  For all curves $m=0.015$.  Solid line: ${\cal C}=\bar{\cal C}=0.51$; dash-dot-dot line: ${\cal C}=1/4$; dotted line: ${\cal C}=0$, namely, the bare vertex, which defines the rainbow truncation; and dash-dot curve: ${\cal C}=-1/8$, for comparison with Ref.\,\protect\cite{detmold}.  In addition, for ${\cal C}=0.51$: dash-dash-dot line - $\alpha_1(s)$ obtained with only the $i=0,1$ terms retained in Eq.\,(\protect\ref{vtxsummeda}), which is the one-loop corrected vertex; and short-dash line - two-loop-corrected vertex.  Dimensioned quantities are measured in units of ${\cal G}$ in Eq.\,(\protect\ref{mnmodel}).}
\end{figure} 

In the upper panel of Fig.\,\ref{a1s} we illustrate the current-quark-mass-dependence of the scalar function associated with $\gamma_\mu$ in the dressed-quark-gluon vertex.  In QCD this is the only function that displays an ultraviolet divergence at one-loop in perturbation theory.  As with $A(p^2)$, $\alpha_1^{\cal C}(p^2)$ evolves slowly with the current-quark mass but again, when the current-quark mass becomes commensurate with or exceeds the theory's mass-scale, it acts to very effectively dampen the momentum dependence of this function so that  $\alpha_1^{\cal C}(p^2)\approx 1$.  This effect is apparent in the rainbow vertex model employed in Ref.\,\cite{mandar} to describe quenched-QCD lattice data.

The lower panel of Fig.\,\ref{a1s} portrays the ${\cal C}$-dependence of $\alpha_1^{\cal C}(p^2)$.  It is here particularly useful to employ the rainbow-ladder result, ${\cal C}=0$, illustrated with the dotted line, as our reference point, because this makes the contrast between the effect of attraction and repulsion in the colour-octet quark-antiquark scattering kernel abundantly clear.  Attraction uniformly increases the magnitude of $\alpha_1^{\cal C}(p^2)$, while the opposite outcome is produced by omitting the effect of the three-gluon-vertex in the DSE for the dressed-quark-gluon vertex.  

\begin{figure}[t] 
\vspace*{2em}
 
\centerline{\includegraphics[width=0.45\textwidth]{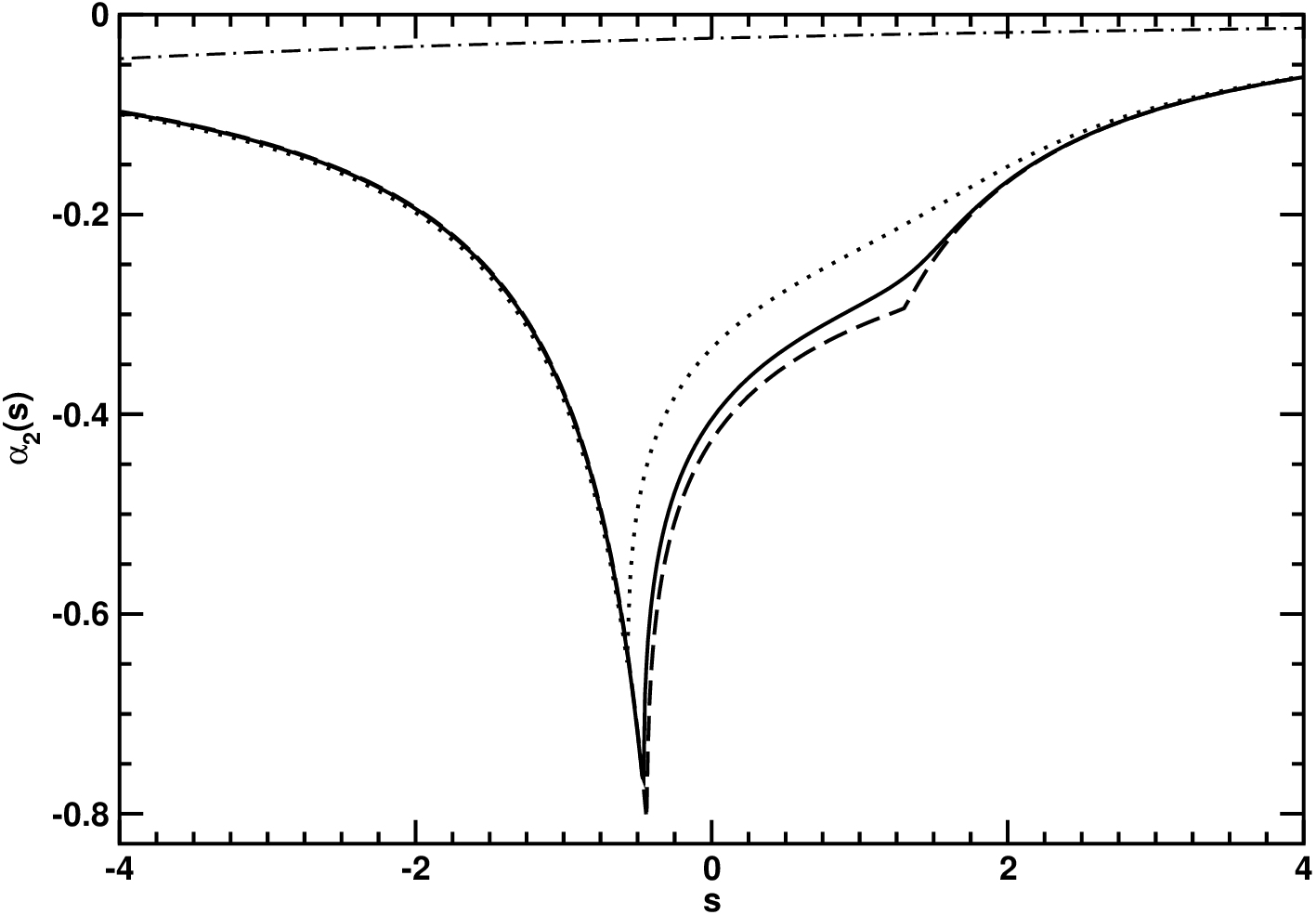}}

\vspace*{3.5em}

\centerline{\includegraphics[width=0.46\textwidth]{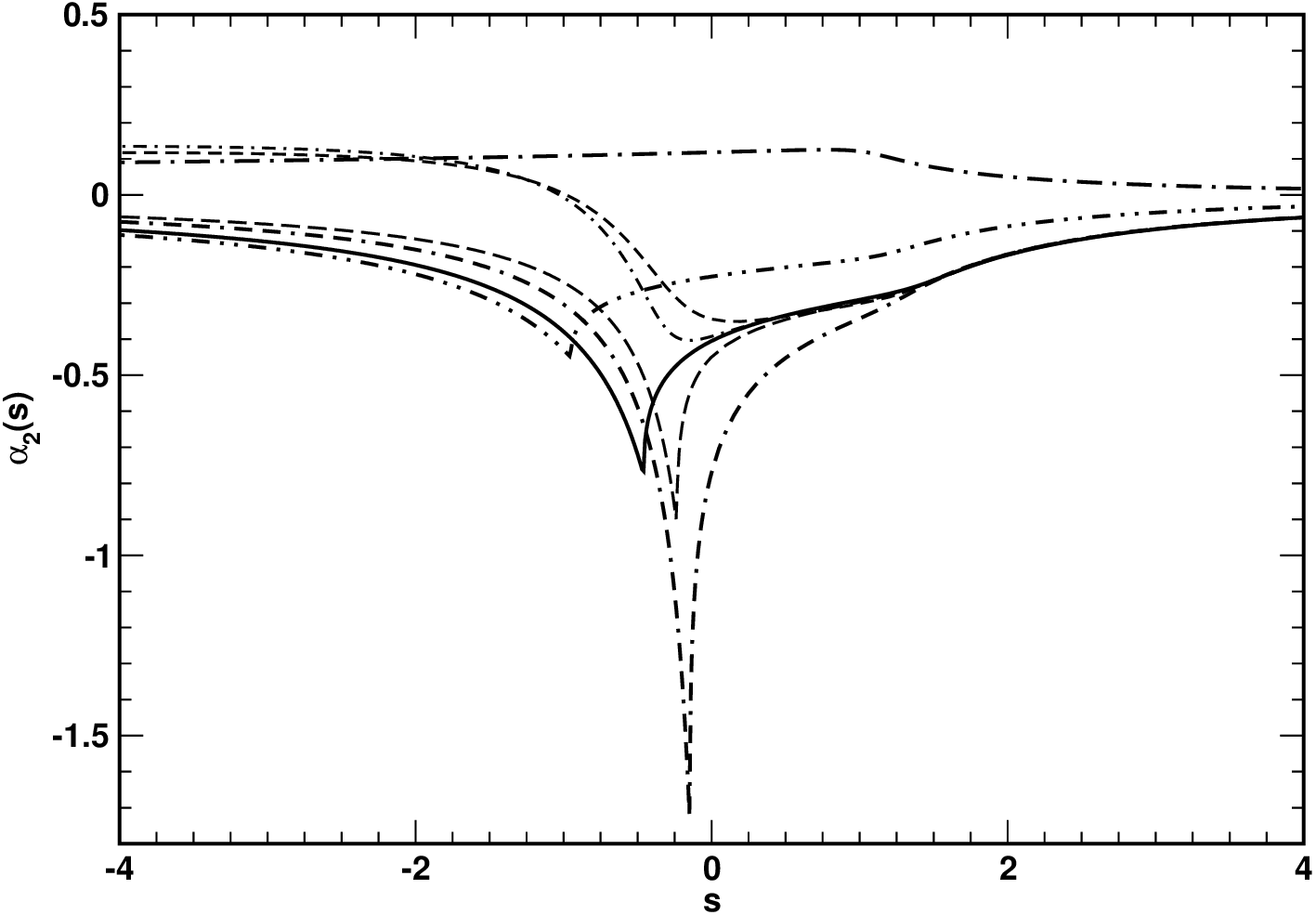}}
 
\caption{\label{a2s} \textit{Upper panel} -- Current-quark-mass-dependence of $\alpha_2^{\cal C}(s)$ in Eq.\,(\protect\ref{GCa123}).  For all curves ${\cal C} = \bar{\cal C} = 0.51$.  Dash-dot line: $m=2$; dotted line: $m=m_{60}$; solid line: $m=0.015$; dashed line: chiral limit, $m=0$.  \textit{Lower panel} -- ${\cal C}$-dependence of this function.  For all curves $m=0.015$.  Solid line: ${\cal C}=\bar{\cal C}=0.51$; dash-dot-dot line: ${\cal C}=1/4$; and dash-dot curve: ${\cal C}=-1/8$, for comparison with Ref.\,\protect\cite{detmold}.  Moreover, for ${\cal C}=0.51$: dash-dash-dot line - one-loop result for $\alpha_2^{\cal C}(s)$, namely, the result obtained with only the $i=0,1$ terms retained in Eq.\,(\protect\ref{vtxsummeda}); short-dash line - two-loop-corrected result; long-dash line - three-loop-corrected; and short-dash-dot line: four-loop corrected.  NB.\, There is no ${\cal C}=0$ curve because $\alpha_2^{\cal C}(s)\equiv 0$ in the bare vertex.  Dimensioned quantities are measured in units of ${\cal G}$ in Eq.\,(\protect\ref{mnmodel}).}
\end{figure} 

The upper panel of Fig.\,\ref{a2s} illustrates the current-quark-mass-dependence of the scalar function modulating the subleading Dirac vector component of the dressed-quark-gluon vertex.  For light-quarks this function is material in size and negative in magnitude.  However, its importance, too, diminishes with increasing current-quark mass.  In the lower panel we show the ${\cal C}$-dependence of $\alpha_2^{\cal C}(p^2)$.  The qualitative features of the completely resummed result for $\alpha_1^{\cal C}(p^2)$ are also manifest here.  However, for this component of the vertex, which is purely dynamical in origin, there is a marked difference at timelike momenta between the result obtained with an odd number of loop corrections in the vertex and that obtained with an even number.  Here the dual pathway pointwise convergence to the fully resummed result is conspicuous.  

It is notable that the size of our complete result for $\alpha_2^{\cal C}(p^2)$ is an order of magnitude smaller than that reported in Ref.\,\cite{skullerud}.  This is an isolated case, however.  The calculated magnitudes of the other functions in the dressed-quark propagator and dressed-quark-gluon vertex are commensurate with those obtained in  quenched-QCD lattice simulations.  We remark in addition that the lattice result is an order of magnitude larger than that obtained with a commonly used vertex \textit{Ansatz} \cite{bc80}.  This curious discrepancy deserves study in more realistic models.

\begin{figure}[t] 
\vspace*{2em}
 
\centerline{\includegraphics[width=0.45\textwidth]{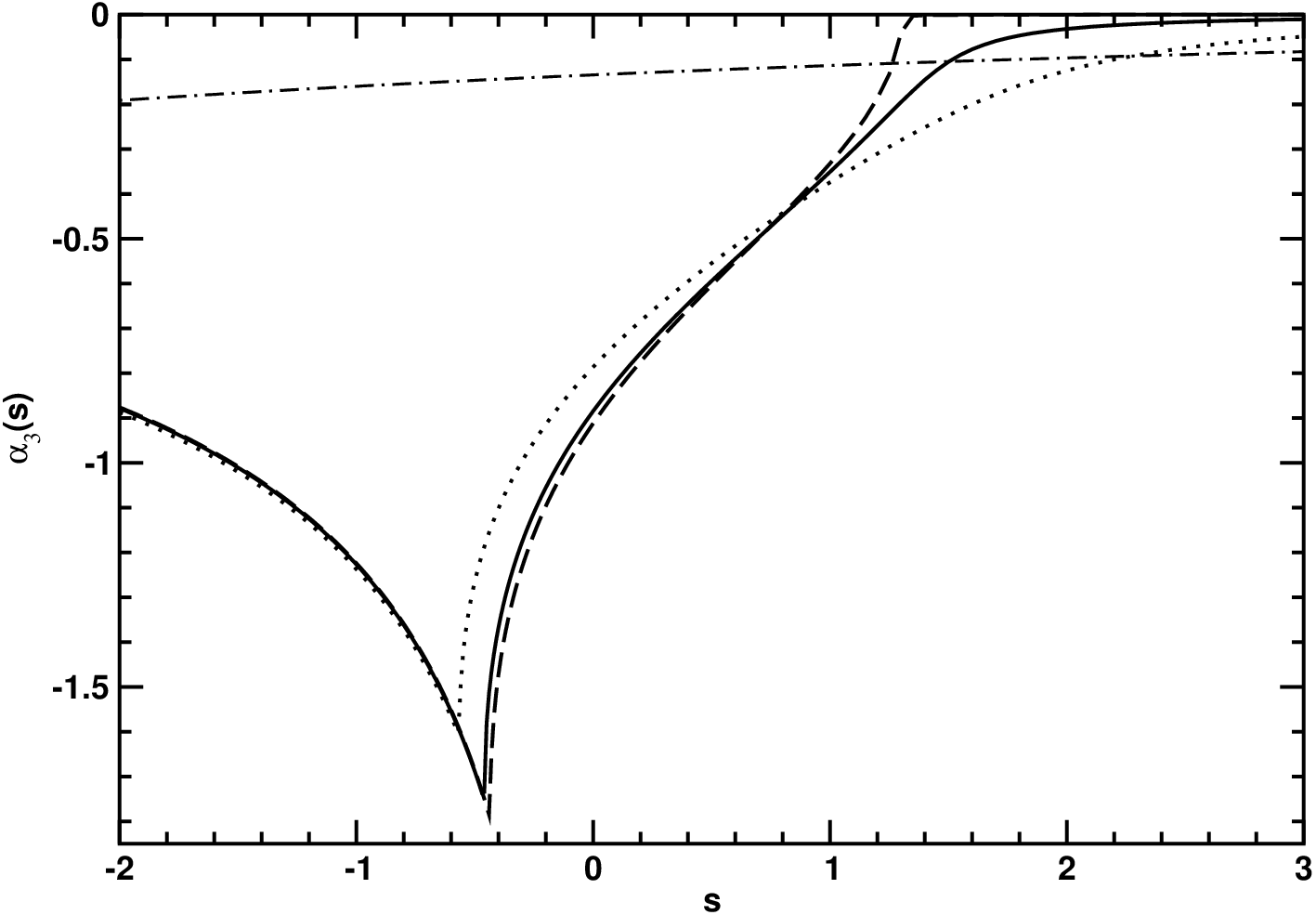}}

\vspace*{3.5em}

\centerline{\includegraphics[width=0.46\textwidth]{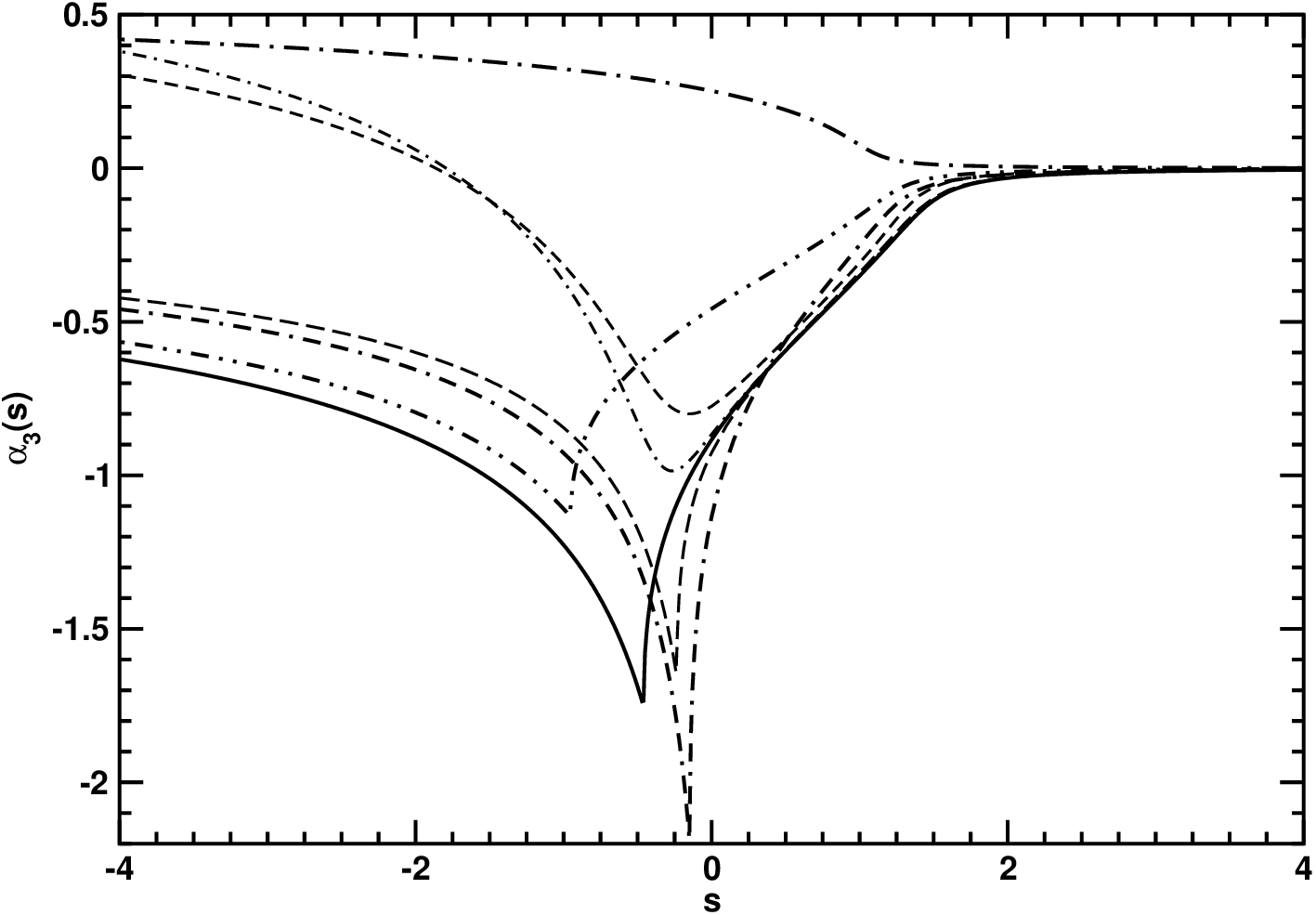}}
 
\caption{\label{a3s} \textit{Upper panel} -- Current-quark-mass-dependence of $\alpha_3^{\cal C}(s)$ in Eq.\,(\protect\ref{GCa123}).  For all curves ${\cal C} = \bar{\cal C} = 0.51$.  Dash-dot line: $m=2$; dotted line: $m=m_{60}$; solid line: $m=0.015$; dashed line: chiral limit, $m=0$.  \textit{Lower panel} -- ${\cal C}$-dependence of this function.  For all curves $m=0.015$.  Solid line: ${\cal C}=\bar{\cal C}=0.51$; dash-dot-dot line: ${\cal C}=1/4$; and dash-dot curve: ${\cal C}=-1/8$, for comparison with Ref.\,\protect\cite{detmold}.  Moreover, for ${\cal C}=0.51$: dash-dash-dot line - one-loop result for $\alpha_3^{\cal C}(s)$, namely, the result obtained with only the $i=0,1$ terms retained in Eq.\,(\protect\ref{vtxsummeda}); short-dash line - two-loop-corrected result; long-dash line - three-loop-corrected; and short-dash-dot line: four-loop corrected.  NB.\, There is no ${\cal C}=0$ curve because $\alpha_3^{\cal C}(s)\equiv 0$ in the bare vertex.  Dimensioned quantities are measured in units of ${\cal G}$ in Eq.\,(\protect\ref{mnmodel}).}
\end{figure} 

In Fig.\,\ref{a3s} we display the result for what might be called the scalar part of the dressed-quark-gluon vertex; viz., $\alpha_3^{\cal C}(p^2)$.  This appellation comes from the feature, apparent in Eq.\,(\ref{a3mnuv}), that $\alpha_3^{\cal C}(p^2)$ is the piece of the vertex whose ultraviolet behaviour is most sensitive to the current-quark mass.  The figure demonstrates, too, that at infrared momenta $\alpha_3^{\cal C}(p^2)$ is materially affected by the scale of DCSB, Eq.\,(\ref{qbqCbar}): at $s=0$ the deviation from its  value in the rainbow truncation is approximately four times that exhibited by $\alpha_1^{\cal C}(p^2)$.  Hence, this term can be important at infrared and intermediate momenta.  However, for light quarks, owing to its sensitivity to the current-quark-mass, it rapidly becomes unimportant at ultraviolet momenta.  The dual pathway pointwise convergence to the fully resummed result in the timelike region is also prominent here.  

\begin{figure}[t] 
\vspace*{2em}
 
\centerline{\includegraphics[width=0.45\textwidth]{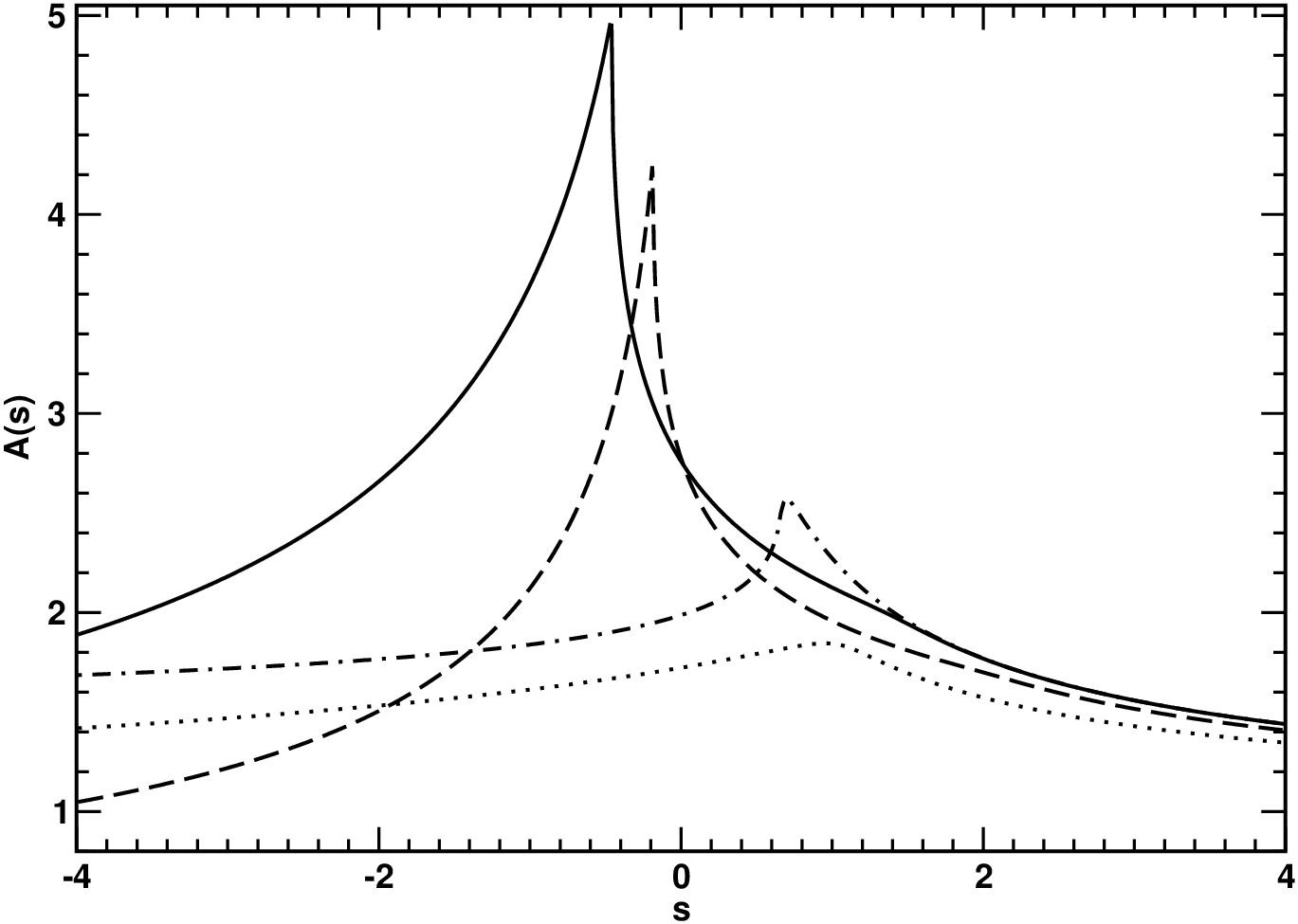}}

\vspace*{3.5em}

\centerline{\includegraphics[width=0.46\textwidth]{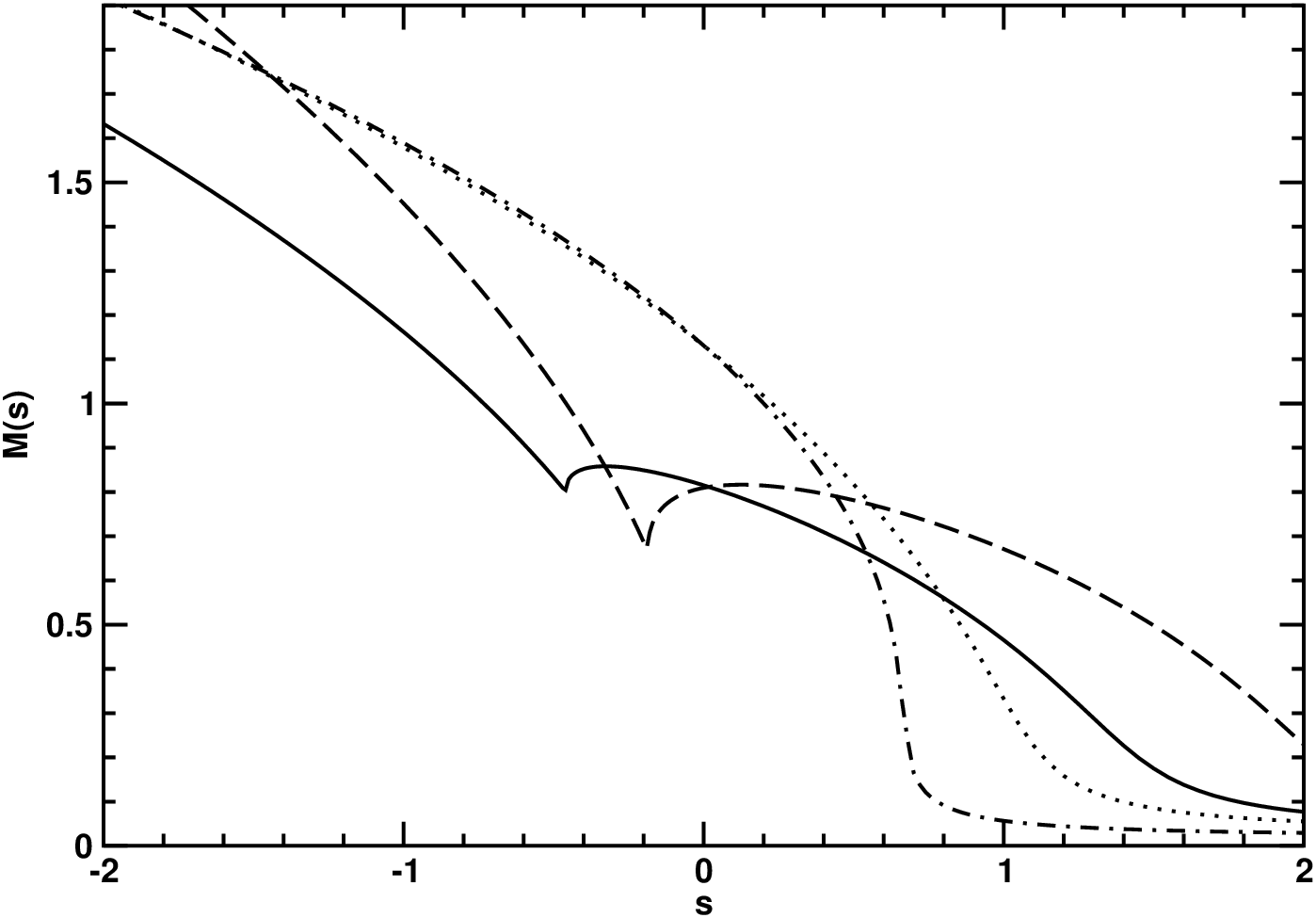}}
 
\caption{\label{AM23} \textit{Upper panel} --  Impact of $\alpha_2^{\cal C}(s)$ and $\alpha_3^{\cal C}(s)$ on $A(s)$.  For ${\cal C} = \bar{\cal C}$, Eq.\,(\ref{Cbar}), solid line: result obtained with both terms present in the vertex; dashed-line: $\alpha_2^{\cal C}(s)$ omitted; dash-dot line: $\alpha_3^{\cal C}(s)$ omitted.  The dotted line is the result obtained with both terms present in the vertex but ${\cal C} = -1/8$.  
\textit{Lower panel} -- Impact of $\alpha_2^{\cal C}(s)$ and $\alpha_3^{\cal C}(s)$ on $M(s)$.  The legend is identical.  In all cases $m=0.015$ and dimensioned quantities are measured in units of ${\cal G}$ in Eq.\,(\protect\ref{mnmodel}).}
\end{figure}

Finally, since they are absent in rainbow truncation, it is illuminating to unfold the different roles played by $\alpha_2^{\cal C}(s)$ and $\alpha_3^{\cal C}(s)$ in determining the behaviour of the self-consistent solution of the gap equation.  Some of these effects are made evident in   Fig.\,\ref{AM23} and all can readily be understood because Eqs.\,(\ref{Asfull}) and (\ref{Bsfull}) signal much of what to expect.  The key observation is that $\alpha_3^{\cal C}(s)$ alone is the source of all coupling between Eqs.\,(\ref{Asfull}) and Eq.\,(\ref{Bsfull}) that is not already present in rainbow-ladder truncation: $\alpha_3^{\cal C}\,B$ appears in the equation for $A(s)$ and $\alpha_3^{\cal C}\,s\,A$ appears in the equation for $B(s)$.  The action of $\alpha_2^{\cal C}$ is merely to modify the rainbow-ladder coupling strengths. 

Turning to detail, each equation involves the combination $s \,\alpha_2^{\cal C}(s)$, which indicates that the impact of $\alpha_2^{\cal C}(s)$ on both $A(s)$ and $B(s)$ is negligible at infrared momenta but possibly material on the intermediate spacelike domain.  This suppression is also felt by $\alpha_3^{\cal C}(s)$ in Eq.\,(\ref{Bsfull}) but not in Eq.\,(\ref{Asfull}).  Hence, omitting $\alpha_3^{\cal C}(s)$ will affect $A(s)$ at infrared momenta but not $B(s)$.  These expectations are easily substantiated by repeating the analysis that gave Eqs.\,(\ref{a00m}), (\ref{b00m}).  When $\alpha_3^{\cal C}\equiv 0$ is imposed in the gap equation one finds $a_0^0 = 2$; namely, the same result that is obtained in the rainbow truncation, but no change in $b_0^0$.  For ${\cal C}>0$ this naturally leads to an increase in $M(0)= b_0^0/a_0^0$.  On the other hand, fixing $\alpha_2^{\cal C}(s)\equiv 0$ produces no change in $a_0^0$ or $b_0^0$.  These results are apparent in the figure.  (NB.\ It is a numerical accident that the dash-dot and dotted curves in the lower panel almost coincide at $s=0$.)

It will readily be appreciated that neither $\alpha_2^{\cal C}(s)$  nor $\alpha_3^{\cal C}(s)$ can affect the deep ultraviolet behaviour of the gap equation's solution, Eqs.\,(\ref{Auv}), (\ref{Buv}), (\ref{ABUV}), because they vanish too rapidly as $1/s \to 0$.  This is plain in Fig.\,\ref{AM23}.

The effects at intermediate spacelike momenta can also be inferred  by inspection of Eqs.\,(\ref{Asfull}), (\ref{Bsfull}).  Both $\alpha_2^{\cal C}(s)$ and $\alpha_3^{\cal C}(s)$ are negative and hence serve to magnify $A(s)$ with respect to the value obtained using a bare vertex.  However, they compete in Eq.\,(\ref{Bsfull}), with $\alpha_2^{\cal C}(s)$ working to diminish $B(s)$ and $\alpha_3^{\cal C}(s)$ acting to amplify it.  In the absence of $\alpha_3^{\cal C}(s)$, therefore, one should expect $M(s)= B(s)/A(s)$ to be suppressed at intermediate momenta because the numerator is reduced and the denominator increased, and consequently a condensate much reduced in magnitude.  Omitting $\alpha_2^{\cal C}(s)$ should yield the opposite effect.  This is precisely the outcome of our numerical studies: 
\begin{eqnarray}
\nonumber
\left.\frac{\langle \bar q q \rangle^0_{{\cal C}=0}}
{\langle \bar q q \rangle^0_{{\cal C}=\bar{\cal C}}}
\right|_{\alpha_3^{\cal C}\equiv 0} = 1.97\,; & \, {\rm and}\, & 
\left.\frac{\langle \bar q q \rangle^0_{{\cal C}=0}}
{\langle \bar q q \rangle^0_{{\cal C}=\bar{\cal C}}}
\right|_{\alpha_2^{\cal C}\equiv 0} = 0.40\,.\\
\label{qbqCbara30}
\end{eqnarray}

At timelike momenta the omission of $\alpha_2^{\cal C}(s)$ has little qualitative impact on $A(s)$ and $M(s)$ because, as we have seen, $\alpha_3^{\cal C}(s)$ is always the dominant term at infrared momenta and hence provides the primary seed for evolution to $s<0$.  In the absence of $\alpha_3^{\cal C}(s)$, on the other hand, the momentum dependence of $A(s)$ and $M(s)$ must resemble that obtained in rainbow truncation because $s \alpha_2^{\cal C}(s)$ has only a small domain of material support.  Figure \ref{AM23} makes this clear. (NB.\ The near pointwise agreement of the dash-dot and dotted curves in the lower panel is a numerical accident.) 
Importantly, the confining character of the dressed-quark propagator is preserved in both cases.  Indeed, omitting both $\alpha_2^{\cal C}(s)$ and $\alpha_3^{\cal C}(s)$ simultaneously does not qualitatively alter this feature of the model prescribed by Eq.\,(\ref{mnmodel}), although the omission does affect aspects of its expression.  That, however, is unsurprising.  It is evident, for example, in a comparison of Refs.\,\cite{munczek86,brw92} with Ref.\,\cite{mn83}. 

\section{Bethe-Salpeter Equation}
\label{secfour} 
The renormalised homogeneous Bethe-Salpeter equation (BSE) for the
quark-antiquark channel denoted by $M$ can be compactly expressed as
\begin{equation} 
\label{bsegen} 
    [\Gamma_M(k;P)]_{EF} =\int_q^\Lambda\! 
    [ K(k,q;P)]_{EF}^{GH}\, [\chi_M(q;P)]_{GH} 
\end{equation} 
where: $k$ is the relative momentum of the quark-antiquark pair and $P$ is
their total momentum; $E, \ldots ,H$ represent colour, flavour and spinor
indices; and
\begin{equation} 
\chi_M(k;P) = S(k_+)\, \Gamma_M(k;P) \,S(k_-)\,, 
\end{equation} 
with $\Gamma_M(q;P)$ the meson's Bethe-Salpeter amplitude.  In
Eq.\,(\ref{bsegen}), $K$ is the fully-amputated dressed-quark-antiquark scattering kernel. 

\subsection{Vertex consistent kernel}
The preservation of Ward-Takahashi identities in those channels related to 
strong interaction observables requires a conspiracy between the 
dressed-quark-gluon vertex and the Bethe-Salpeter 
kernel~\cite{truncscheme,herman}.  The manner in which these constraints are realised for vertices of the class considered herein was made explicit in  Ref.\,\cite{detmold}.  In that systematic and nonperturbative truncation scheme the rainbow gap equation and ladder Bethe-Salpeter equation represent the lowest-order Ward-Takahashi identity preserving pair.  Beyond this, each additional term in the vertex generates a unique collection of terms in the Bethe-Salpeter kernel, a subset of which are always nonplanar. 

For any dressed-quark-gluon vertex appearing in the gap equation, which can be represented expressly by an enumerable series of contributions, the Bethe-Salpeter kernel that guarantees the validity of all Ward-Takahashi identities is realised in 
\begin{eqnarray} 
\nonumber \lefteqn{\Gamma_M(k;P)  =  \int_q^\Lambda 
{\cal D}_{\mu\nu}(k - q)\, l^a \gamma_\mu }\\ 
\nonumber &&  \times \, \left[\rule{0em}{3ex} \chi_M(q;P) \, 
l^a\, 
\Gamma_\nu(q_-,k_-) + S(q_+) \, \Lambda_{M\nu}^{a}(q,k;P)\right], \\ 
\label{genbsenL1}
\end{eqnarray} 
where
\begin{equation} 
\Lambda_{M\nu}^{a}(q,k;P) = \sum_{n=0}^\infty \Lambda_{M\nu}^{a;n}(q,k;P)\,, 
\label{Lambdatotal} 
\end{equation} 
with 
\begin{eqnarray} 
\nonumber \lefteqn{-\frac{1}{8 \, {\cal C}}\,
\Lambda^{a;n}_{M\nu}(\ell,k;P) = \int_q^\Lambda\! {\cal 
D}_{\rho\sigma}(\ell-q)\,l^b \gamma_\rho\, \chi_M(q;P)\,  }\\
\nonumber
&& \times \,
l^a \Gamma_{\nu, {n-1}}^{\cal C}(q_-,q_-+k-\ell) 
\, S(q_-+k-\ell)\, l^b \gamma_\sigma\\ 
\nonumber &+& \int_{q}^\Lambda\! {\cal D}_{\rho\sigma}(k -q)\, l^b
\gamma_\rho\,  
S(q_+ + \ell - k)\, \\
\nonumber
&& \times \, l^a \Gamma_{\nu,n-1}^{\cal C}(q_+ + \ell - k,q_+)\, \chi_M(q;P) 
\,l^b \gamma_\sigma \\
&+&
\nonumber
 \int_{q^\prime}^\Lambda\! {\cal D}_{\rho\sigma}(\ell -q^\prime) l^b 
\gamma_\rho\, S(q^\prime_+)\, \\
\nonumber
&& \times \, \Lambda^{a;n-1}_\nu(q^\prime,q^\prime 
+k - \ell;P)\, S(q_-^\prime+k-\ell) \, l^b \gamma_\sigma . \\ \label{Lambdarecursion}  
\end{eqnarray} 
This last equation is a recursion relation, which is to be solved subject to the initial condition $\Lambda_{M \nu}^{a;0}\equiv 0 $.

It is apparent that the kernel of Eq.\,(\ref{genbsenL1}) is completely known  once the vertex to be used in the gap equation is fixed and that equation's solution is obtained.

The Bethe-Salpeter amplitude for any meson can be written in the form
\begin{equation} 
\label{genbsa} \Gamma_M(k;P) = \mbox{\boldmath $I$}_c \sum_{i=1}^{N_M}\, {\cal 
G}^i(k;P) \,f^i_M(k^2,k\cdot P;P^2) =: \left[ \mbox{\boldmath $G$}\right]\, \mbox{\boldmath $f$}_M \, ,
\end{equation} 
where ${\cal G}^i(k;P)$ are those independent Dirac matrices required to span the space containing the meson under consideration.  It then follows upon substitution of this formula that Eq.\,({\ref{Lambdarecursion}) can be written compactly as 
\begin{equation}
\label{Leasy}
\Lambda_{M \nu}^{a;n} = \left\{\left[ {\cal K}_{M \nu}^{i} \, \alpha_{i;n-1}^{\cal C} \right]  + \left[ {\cal L} \, \Lambda _{M \nu }^{a;n-1} \right]\right\} \mbox{\boldmath $f$}_M\,.
\end{equation}
This expression states that $\Lambda_{M \nu}^{a;n}$ can primarily be considered as a matrix operating in the space spanned by the independent components of the Bethe-Salpeter amplitude, with its Dirac and Lorentz structure projected via the contractions in the BSE.  The first term (${\cal K}_M$) in Eq.\,(\ref{Leasy}) represents the contribution from the first two integrals on the right-hand-side (r.h.s.) of Eq.\,({\ref{Lambdarecursion}).  This is the driving term in the recursion relation.  The second term (${\cal L}$) represents the last integral, which enacts the recursion.  

\subsection{Solutions of the vertex-consistent meson Bethe-Salpeter equation} 
A discussion of the general features of the vertex-consistent BSE and its solution is given in Ref.\,\cite{detmold}.  Herein we therefore focus immediately on the model under consideration.

\subsubsection{\mbox{\boldmath $\pi$}-meson}
With the model of the dressed-gluon interaction in Eq.\,(\ref{mnmodel}) the relative momentum between a meson's constituents must vanish.  It follows that the general form of the Bethe-Salpeter amplitude for a pseudoscalar meson of equal-mass constituents is
\begin{equation} 
\label{pimodel}
\Gamma_\pi(P) = \gamma_5\,\left[ i f^\pi_{ 1}(P^2) + \gamma\cdot \hat
P\,f^\pi_{ 2}(P^2)\right],
\end{equation} 
where $\hat P$, $\hat P^2=1$,  is the direction-vector associated with $P$.

To obtain the vertex-consistent BSE one must first determine $\Lambda_{\pi\nu}^{a;1}$.  That is obtained by substituting Eq.\,(\ref{pimodel}) into the r.h.s.\ of Eq.\,({\ref{Lambdarecursion}).  Only the first two integrals contribute because of the initial condition and they are actually algebraic expressions when Eq.\,(\ref{mnmodel}) is used.  This gives ${\cal K}_{\pi \nu}^{i}$ in Eq.\,(\ref{Leasy}).  Explicit calculation shows this to be identically zero, and hence $\Lambda_{\pi\nu}^{a;1} \equiv 0$.  Furthermore because this is the driving term in the recursion relation then
\begin{equation}
\Lambda_{\pi \nu}^{a}(k,k;P) \equiv 0\,.
\end{equation}
The result is not accidental.  It is fundamentally a consequence of the fact that $\Lambda^a_{\pi\mu}(\ell,k;P)= l^a \,\Lambda_{\pi\mu}(\ell,k;P)$ is an axial-vector vertex which transforms under charge conjugation according to:
\begin{equation} 
\bar\Lambda_{\pi\mu}(-k,-\ell;P)^{\rm t} = - \Lambda_{\pi\mu}(\ell,k;P)\,, 
\end{equation} 
where $(\cdot)^{\rm t}$ denotes matrix transpose, and Eq.\,(\ref{mnmodel}) enforces $k=0=\ell$ in the BSE.  It is not a general feature of the vertex-consistent Bethe-Salpeter kernel. 

One thus arrives at a particularly simple vertex-consistent BSE for the pion (Q=P/2):
\begin{equation} 
\Gamma_\pi(P) =- \,\gamma_\mu\, 
S(Q)\,\Gamma_\pi(P)\,S(-Q)\,\Gamma_\mu^{\cal C}(-Q)\,. \label{pibsemnmodel} 
\end{equation} 

Consider the matrices
\begin{equation}
{\cal P}_1 = -\sfrac{i}{4} \,\gamma_5\,,\; 
{\cal P}_2 = \sfrac{1}{4}\, \gamma\cdot \hat P\,\gamma_5\,,
\end{equation} 
which satisfy
\begin{equation}
f^\pi_{i}(P)= {\rm tr}_{\rm D} {\cal P}_i \, \Gamma_\pi(P)\,.
\end{equation}
They may be used to rewrite Eq.\,(\ref{pibsemnmodel}) in the form
\begin{equation}
\label{fHf}
\mbox{\boldmath $f$}_\pi(P) = {\cal H}_\pi(P^2) \mbox{\boldmath $f$}_\pi(P)\,,
\end{equation}
wherein ${\cal H}_\pi(P^2)$ is a $2\times 2$ matrix 
\begin{eqnarray}
\nonumber 
\lefteqn{
{\cal H}_\pi(P^2)_{ij} =}\\
\nonumber 
&& -\frac{\delta }{\delta f^\pi_{ j}}{\rm tr}_{\rm D} {\cal P}_i \, \gamma_\mu\, S(Q)\,\Gamma_\pi(P)\,S(-Q)\,\Gamma_\mu^{\cal C}(-Q)\,.\\
\end{eqnarray}

Equation (\ref{fHf}) is a matrix eigenvalue problem in which the kernel ${\cal H}$ is a function of $P^2$.  This equation has a nontrivial solution if, and only if, at some $M^2$ 
\begin{equation} 
\label{characteristic} \left.\det\left[{\cal H}_\pi(P^2) - \mbox{\boldmath 
$I$}\right]\right|_{P^2+M^2=0} = 0\,.
\end{equation} 
The value of $M$ for which this characteristic equation is satisfied is the bound state's mass.  NB.\ In the absence of a solution there is no bound state in the channel under consideration.
 
\begin{table}[t] 
\caption{\label{tablea} Calculated $\pi$ and $\rho$ meson masses, in GeV, quoted with ${\cal G}=0.69$ GeV, in which case $m=0.015\, {\cal G} =
10\,$MeV.  (In the notation of Ref.~\protect\cite{truncscheme}, this value of
${\cal G}$ corresponds to $\eta = 1.39\,$GeV.) $n$ is the number of
loops retained in dressing the quark-gluon vertex, see Eq.\,(\protect\ref{vtxsummeda}), and hence the order of the vertex-consistent
Bethe-Salpeter kernel. NB.\ $n=0$ corresponds to the rainbow-ladder truncation, in which case $m_\rho = \sqrt{2}\, {\cal G}$, and that is why this column's results are independent of ${\cal C}$.\vspace*{1ex}}
\begin{ruledtabular} 
\begin{tabular*} 
{\hsize} {l@{\extracolsep{0ptplus1fil}} 
l@{\extracolsep{0ptplus1fil}}
|c@{\extracolsep{0ptplus1fil}}c@{\extracolsep{0ptplus1fil}} 
c@{\extracolsep{0ptplus1fil}}c@{\extracolsep{0ptplus1fil}}} 
%
 & & $M_H^{n=0}$ & $M_H^{n=1}$ & $M_H^{n=2}$ & $M_H^{n=\infty}$\\\hline 
\rule{0mm}{3ex}${\cal C}= -(1/8)$ &$\pi$, $m=0$ & 0 & 0 & 0 & 0\\ 
&$\pi$, $m=0.01$ & 0.149 & 0.153 & 0.154 & 0.154\\
&$\rho$, $m=0$ & 0.982 & 1.074 & 1.089 & 1.091\\ 
&$\rho$, $m=0.01$ & 0.997 & 1.088 & 1.103 & 1.105 \\\hline
\rule{0mm}{3ex}${\cal C}= (1/4)$ &$\pi$, $m=0$ & 0 & 0 & 0 & 0\\ 
&$\pi$, $m=0.01$ & 0.149 & 0.140 & 0.142 & 0.142\\
&$\rho$, $m=0$ & 0.982 & 0.789 & 0.855 & 0.842\\ 
&$\rho$, $m=0.01$ & 0.997 & 0.806 & 0.871 & 0.858 \\\hline
\rule{0mm}{3ex}${\cal C}=\bar{\cal C}= 0.51$ &$\pi$, $m=0$ & 0 & 0 & 0 & 0\\ 
&$\pi$, $m=0.01$ & 0.149 & 0.132 & 0.140 & 0.138\\
&$\rho$, $m=0$ & 0.982 & \ldots  & 0.828 & 0.754\\ 
&$\rho$, $m=0.01$ & 0.997 & \ldots & 0.844 & 0.770 \\
\end{tabular*} 
\end{ruledtabular} 
\end{table} 

We have solved Eq.\,(\ref{characteristic}) for the pion and the results are presented in Table \ref{tablea}.  The fact that the vertex-consistent Bethe-Salpeter kernel ensures the preservation of the axial-vector Ward-Takahashi identity, and hence guarantees the pion is a Goldstone boson in the chiral limit, is abundantly clear: irrespective of the value of ${\cal C}$ and the order of the truncation, $m_\pi = 0$ for $m=0$.  Away from this symmetry-constrained point the results indicate that, with net attraction in the colour-octet quark-antiquark scattering kernel, the rainbow-ladder (lowest order) truncation overestimates the mass; i.e., it yields a value greater than that obtained with the fully resummed vertex $(n=\infty)$.  Moreover, the approach to the exact result for the mass is not monotonic.  On the other hand, given two truncations for which solutions exist, characterised by $n_1$- and $n_2$-loop insertions, respectively, then
\begin{equation}
|M_H^{n=\infty} - M_H^{n_2}| < |M_H^{n=\infty} - M_H^{n_1}|\,,\; n_2>n_1\,;
\end{equation}
viz., correcting the vertex improves the accuracy of the mass estimate.

\subsubsection{\mbox{\boldmath $\rho$}-meson}
In our algebraic model the complete form of the Bethe-Salpeter amplitude for a vector meson is 
\begin{equation} 
\Gamma_\rho^\lambda(P) = \gamma \cdot\epsilon^\lambda(P)\, f_1^\rho(P^2) + 
\sigma_{\mu\nu}\,\epsilon_\mu^\lambda(P)\, \hat P_\nu\, f_2^\rho(P^2) \,. 
\label{gammarho} 
\end{equation} 
This expression, which has only two independent functions, is much simpler
than that allowed by a more realistic interaction, wherein there are eight
terms.  Nevertheless, Eq.\,(\ref{gammarho}) retains the amplitudes that are found to be dominant in more sophisticated studies \cite{pieterrho}.  In
Eq.\,(\ref{gammarho}), $\{\epsilon_\mu^\lambda(P);\, \lambda=-1,0,+1\} $ is
the polarisation four-vector:
\begin{equation} 
P\cdot \epsilon^\lambda(P)=0\,,\; \forall \lambda\,;\;\; 
\epsilon^\lambda(P)\cdot\epsilon^{\lambda^\prime}(P) = \delta^{\lambda 
\lambda^\prime}. 
\end{equation} 

The construction of the vertex-consistent BSE for the class of vertices under consideration herein is fully described in Ref.\,\cite{detmold}.  With the model for the dressed-gluon interaction in Eq.\,(\ref{mnmodel}), that procedure yields an equation that can be written
\begin{equation}
\label{fHrho}
\mbox{\boldmath $f$}_\rho(P) = {\cal H}_\rho(P^2) \mbox{\boldmath $f$}_\rho(P)\,,
\end{equation}
wherein ${\cal H}_\rho(P^2)$ is a $2\times 2$ matrix: 
\begin{equation}
\label{Hrho}
{\cal H}_\rho(P^2) = {\cal H}_{\rho \Gamma}(P^2) + {\cal H}_{\rho \Lambda}(P^2)\,.
\end{equation}

In Eq.\,(\ref{Hrho}), ${\cal H}_{\rho \Gamma}(P^2)$ is analogous to the sole term that remains for the pion; viz., 
\begin{eqnarray}
\nonumber 
\lefteqn{
{\cal H}_{\rho \Gamma}(P^2)_{ij} =}\\
&& -\frac{\delta }{\delta f^\rho_{ j}}{\rm tr}_{\rm D} {\cal P}^{\lambda}_i \,\gamma_\mu\, 
S(Q)\,\Gamma_\rho^\lambda(P)\,S(-Q)\,\Gamma_\mu^{\cal C}(-Q)\,, \label{HGammarho}
\end{eqnarray}
where the projection matrices are
\begin{equation}
{\cal P}_1^{\lambda}= \sfrac{1}{12} \,\gamma\cdot \epsilon^\lambda(P)\,,\; 
{\cal P}_2^\lambda= \sfrac{1}{12} \, 
\sigma_{\mu\nu}\,\epsilon_\mu^\lambda(P)\, \hat P_\nu \,.
\end{equation} 

The remaining term in Eq.\,(\ref{Hrho}) is generated by $\Lambda_{\rho\nu}^a$, which is nonzero in this case, and produces planar and nonplanar contributions.  To be explicit
\begin{eqnarray} 
\Lambda_{\rho\,\nu}^{a}(\epsilon^\lambda,P) & = & l^a\, 
\Lambda_{\rho\,\nu}(\epsilon^\lambda,P)\,, \label{Lambdarhoa} 
\end{eqnarray} 
where the Dirac structure is completely expressed through  
\begin{eqnarray} 
\nonumber \lefteqn{ \Lambda_{\rho\,\nu}(\epsilon^\lambda,P)  = \beta_1(P) 
\epsilon_\nu^\lambda + \beta_2(P)\,\epsilon_\nu^\lambda\,i\gamma\cdot\hat P}\\ %
\nonumber &&  +\, \beta_3(P)\, i\gamma\cdot \epsilon^\lambda\,\hat P_\nu\, + 
\beta_4(P)\,\sigma_{\alpha\beta}\,\epsilon_\alpha^\lambda(P)\, \hat 
P_\beta\,\gamma_\nu\\ 
&& + \,\beta_5(P)\, i\sigma_{\alpha\beta}\,\epsilon_\alpha^\lambda(P)\, \hat 
P_\beta\,\hat P_\nu + \beta_6(P)\, \gamma\cdot \epsilon^\lambda\,\gamma_\nu\,. \label{Lambdarho} 
\end{eqnarray} 
The manner by which the vector $\mbox{\boldmath $\beta$}(P)={\rm column}[\beta_1,\beta_2,\beta_3,\beta_4,\beta_5,\beta_6]$ is determined was described in Ref.\,\cite{detmold} and in our case that procedure yields
\begin{equation}
\mbox{\boldmath $\beta$}(P) = \frac{1}{1 - \mbox{\boldmath $L$}_\rho(P)}\, \mbox{\boldmath $G$}_\rho(P)\,,
\end{equation}
with
\begin{eqnarray} 
\nonumber \lefteqn{ -\frac{1}{2\, {\cal C}}\,\Delta^2\,\mbox{\boldmath 
$L$}_\rho(P) = 
}\\ 
\nonumber & & 
\left[ \begin{array}{cccccc} 
2\,\Delta& 0 & 0 & 0 & 0 &  2\,\Delta\\ 
0& -\,\Delta & 0 & -2\,\Delta & 0 & 0 \\ 
0& 0 & 2 [Q^2 A^2 - B^2] & 2 B^2 & -2 Q A B 
& -2 Q A B \\ 
0& 0 & 0 & \Delta & 0 & 0 \\ 
0 & 0 & 0& 0 & 0 & 0\\ 
0 & 0 & 0& 0 & 0 & 0 \\ 
\end{array} \right]\\ 
&& 
\end{eqnarray} 
in which the argument of each function is $Q^2$, and 
\begin{eqnarray} 
\nonumber \lefteqn{\mbox{\boldmath 
$G$}_\rho(P) = 
}\\ 
\nonumber && 
\frac{- 4 \, {\cal C}\,\alpha_1^{\cal C}(Q)}{\Delta(Q)}
\left[ \begin{array}{c} 
2[ f_1^\rho(P) B(Q^2) + f_2^\rho\,Q A(Q^2)] \\ 
f_1^\rho(P)\,Q\,A(Q^2) -f_2^\rho(P)\,B(Q^2)  \\ 
- f_1^\rho(P)\,Q\,A(Q^2) +f_2^\rho(P)\,B(Q^2)\\ 
- f_1^\rho(P)\,Q\,A(Q^2) +f_2^\rho(P)\,B(Q^2)\\ 
0 \\ 
0  \\ 
\end{array} \right].\\ 
&& 
\end{eqnarray} 
It is correct that in this study $\mbox{\boldmath $G$}_\rho(P)$ depends only on $\alpha_1^{\cal C}$ and, moreover, the zeros in its last two rows entail $\beta_5(P)=0=\beta_6(P)$.

The last term in Eq.\,(\ref{Hrho}) is thus 
\begin{eqnarray}
{\cal H}_{\rho \Lambda}(P^2) &=& 
 -\frac{\delta }{\delta f^\rho_{ j}}{\rm tr}_{\rm D} {\cal P}^{\lambda}_i \,\gamma_\mu\, S(Q)\,\Lambda_{\rho \mu}(P) \label{Hrholast}
\end{eqnarray}
and hence the $\rho$-meson's mass, $M_\rho$, is obtained as the solution of
\begin{eqnarray}
\label{rhochar}
0 &= & \left.\det\left[{\cal H}_\rho(P^2) - \mbox{\boldmath 
$I$}\right]\right|_{P^2+M_\rho^2=0} \\
\nonumber &=& 
\left.\det\left[{\cal H}_{\rho\Gamma}(P^2) + {\cal H}_{\rho\Lambda}(P^2) - \mbox{\boldmath  $I$}\right]\right|_{P^2+M_\rho^2=0}.
\end{eqnarray}

We have solved Eq.\,(\ref{rhochar}) and present the results in Table \ref{tablea}.  One sees that with increasing net attraction in the colour-octet projection of the quark-antiquark scattering kernel the amount by which the rainbow-ladder truncation overestimates the exact mass also increases: with the amount of attraction suggested by lattice data the $n=0$ mass is 27\% too large.  A related observation is that the bound state's mass decreases as the amount of attraction between its constituents increases.  This is unsurprising.  Furthermore, with increasing attraction, even though the fully resummed vertex and consistent kernel always yield a solution, there is no guarantee that a given truncated system will support a bound state: the one-loop corrected vertex and consistent kernel ($n=1$) do not have sufficient binding to support a $\rho$-meson.  This is overcome at the next order of truncation, which yields a mass 9.7\% too large.  Given the results for the dressed-quark propagator and dressed-quark-gluon vertex illustrated in Figs.\,\ref{As} -- \ref{a3s}; namely, the dual pathways of pointwise convergence to the true solutions, the difference between odd- and even-$n$ truncations in our model is understandable.  Nevertheless, the primary observation; viz., a given beyond-rainbow-ladder truncation may not support a bound state even though one is present in the solution of the complete and consistent system, provides a valuable and salutary tip for model building and hadron phenomenology.  

Finally, as has often been observed, and independent of the truncation, bound state solutions of gap-equation-consistent BSEs always 
yield the full amount of $\pi$-$\rho$ mass splitting, even in the chiral limit.  This splitting is plainly driven by the DCSB mechanism and therefore its true understanding requires a veracious realisation of that phenomenon. 

\subsubsection{Dependence on the current-quark mass}
In connection with this last observation it is relevant to explore the evolution with current-quark mass of the pseudoscalar and vector meson masses and, consequently, of the difference between them.  The results for pseudoscalar mesons should be interpreted with the following caveat in mind.  In constructing the vertex and kernel we omitted contributions from gluon vacuum polarisation diagrams.  These contribute only to flavour singlet meson channels.  Hence, for light-quarks in the pseudoscalar channel, wherewith such effects may be important \cite{mecke,klabucar,scoccola}, our results should be understood to apply only to flavour nonsinglets.  In principle, the same is true for light vector mesons.  However, experimentally, the $\omega$ and $\phi$ mesons are almost ideally mixed; i.e., the $\omega$ exhibits no $\bar s s$ content whereas the $\phi$ is composed almost entirely of this combination.  We therefore assume that the vacuum polarisation diagrams we have omitted are immaterial in the study of vector mesons.  (NB.\ It is an artefact of Eq.\,(\ref{mnmodel}) that this model supports neither scalar nor axial-vector meson bound states \cite{mn83,detmold}.)  

\begin{table}[tb] 
\caption{\label{currentquark} Current-quark masses required to reproduce the experimental masses of the vector mesons.   The values of $m_{\eta_c}$, $m_{\eta_b}$ are predictions.  Experimentally \protect\cite{pdg}, $m_{\eta_c}=2.9797\pm 0.00015\,$ and $m_{\eta_b}=9.30\pm 0.03$.  NB.\ $0^-_{s\bar s}$ is a fictitious pseudoscalar meson composed of unlike-flavour quarks with mass $m_s$, which is included for comparison with other nonperturbative studies.  All masses are listed in GeV.\vspace*{1ex}}
\begin{ruledtabular} 
\begin{tabular*} 
{\hsize} {r@{\extracolsep{0ptplus1fil}} 
r@{\extracolsep{0ptplus1fil}}
r@{\extracolsep{0ptplus1fil}}r@{\extracolsep{0ptplus1fil}}} 
%
$m_{u,d}=$ 0.01 & $m_s=$ 0.166 & $m_c =$ 1.33  & $m_b=$ 4.62 \\
$m_\rho = 0.77$ & $m_\phi = 1.02$ & $m_{J/\psi} = 3.10$ & $m_{\Upsilon(1S)}=9.46$\\
$m_\pi = 0.14$ & $m_{0^-_{s\bar s}} = 0.63$ & $m_{\eta_c}=2.97$ & $m_{\eta_b}=9.42$
\end{tabular*} 
\end{ruledtabular} 
\end{table} 

We fix the model's current-quark masses via a fit to the masses of the vector mesons, and the results of this exercise are presented in Table \ref{currentquark}.  The model we're employing is ultraviolet finite and hence our current-quark masses cannot be directly compared with any current-quark mass-scale in QCD.  Nevertheless, the values are quantitatively consistent with the pattern of flavour-dependence in the explicit chiral symmetry breaking mass-scales of QCD.

\begin{figure}[t] 
\vspace*{2em}
 
\centerline{\includegraphics[width=0.45\textwidth]{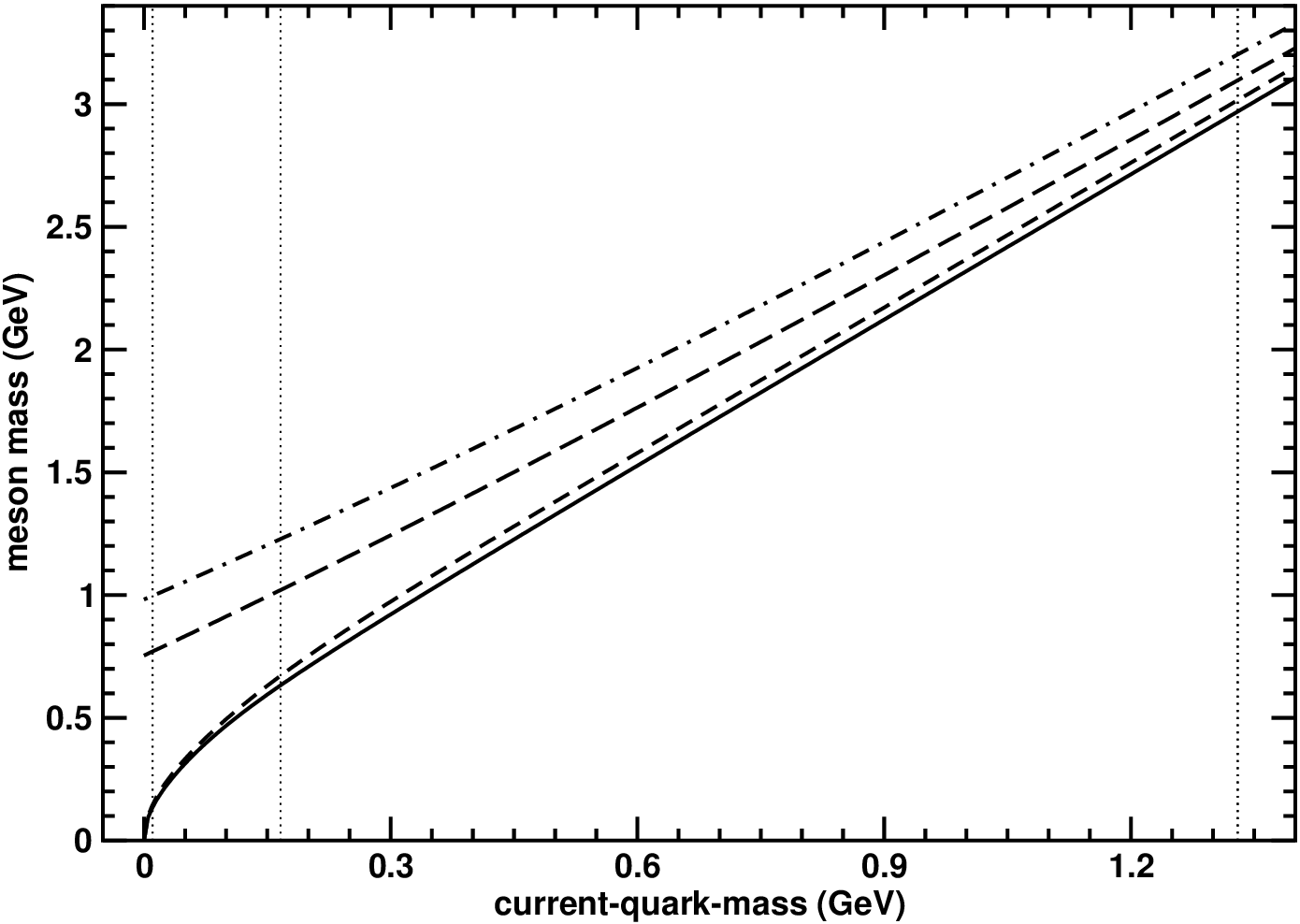}}

\vspace*{3.5em}

\centerline{\includegraphics[width=0.46\textwidth]{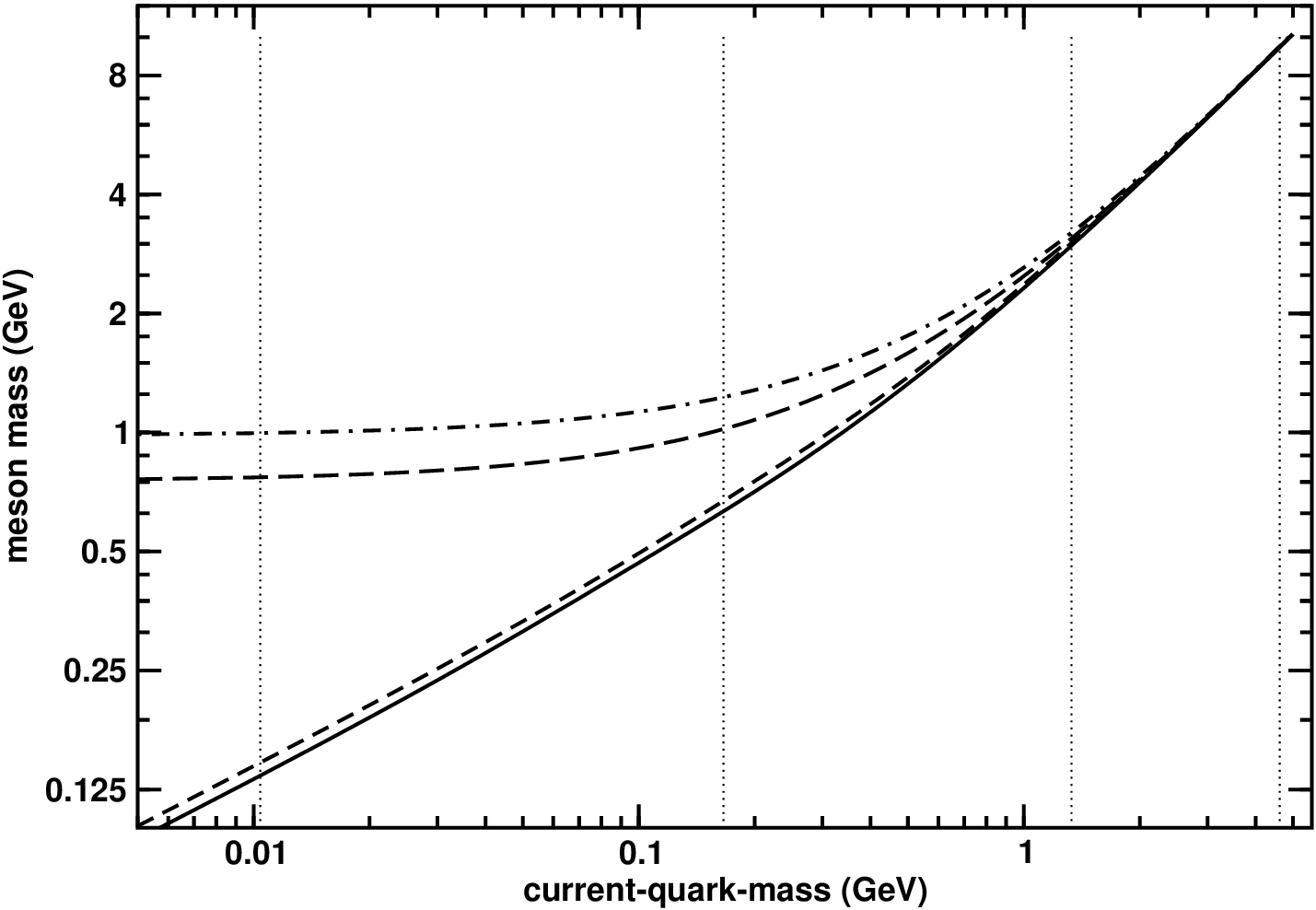}}
 
\caption{\label{massplot} Evolution of pseudoscalar and vector $q \bar q$ meson masses with the current-quark mass.  Solid line: pseudoscalar meson trajectory obtained with ${\cal C}=\bar{\cal C}=0.51$, Eq.\,(\protect\ref{Cbar}), using the completely resummed dressed-quark-gluon vertex in the gap equation and the vertex-consistent Bethe-Salpeter kernel; short-dash line: this trajectory calculated in rainbow-ladder truncation.  Long-dash line: vector meson trajectory obtained with $\bar{\cal C}$ using the completely resummed vertex and the consistent Bethe-Salpeter kernel; dash-dot line: rainbow-ladder truncation result for this trajectory.  The dotted vertical lines mark the current-quark masses in Table \protect\ref{currentquark}.}
\end{figure} 

Our calculated results for the current-quark mass-dependence of pseudoscalar and vector meson masses are presented in Fig.\,\ref{massplot}.
In the neighbourhood of the chiral limit the vector meson mass is approximately independent of the current-quark mass whereas the pseudoscalar meson mass increases rapidly, according to (in GeV)
\begin{equation}
\label{gmor}
m_{0^-}^2 \approx 1.33\, m\,\,\; m\ll {\cal G}\,,
\end{equation}
thereby reproducing the pattern predicted by QCD \cite{mrt98}.

With the model's value of the vacuum quark condensate, Eq.\,(\ref{qbqCbar}), this result allows one to infer the chiral-limit value of $f_\pi^0= 0.056\,$GeV via the Gell-Mann--Oakes--Renner relation.  It is an artefact of the model that the relative-momentum-dependence of Bethe-Salpeter amplitudes is described by $\delta^4(p)$ and so a direct calculation of this quantity is not realistic.  The value is low, as that of the condensate is low, because the model is ultraviolet finite.  In QCD the condensate and leptonic decay constant are much influenced by the high-momentum tails of the dressed-quark propagator and Bethe-Salpeter amplitudes.

The curvature in the pseudoscalar trajectory persists over a significant domain of current-quark mass.  For example, consider two pseudoscalar mesons, one composed of unlike-flavour quarks each with mass $2 m_s$ and another composed of such quarks with mass $m_s$.  In this case
\begin{equation}
\frac{m^2_{0^-_{2 m_s}}}{m^2_{0^-_{m_s}}} = 2.4\,,
\end{equation}
a result which indicates that the nonlinear evolution exhibited in Eq.\,(\ref{gmor}) is still very much in evidence for current-quark masses as large as twice that of the $s$-quark.  With this result our simple model reproduces a feature of more sophisticated DSE studies \cite{mariswien,pcterice,cdrLC01} and a numerical simulation of quenched lattice-QCD \cite{latticeVPS}.

The mode of behaviour just described is overwhelmed when the current-quark mass becomes large: $m\gg {\cal G}$.  In this limit the vector and pseudoscalar mesons become degenerate, with the mass of the ground state pseudoscalar meson rising monotonically to meet that of the vector meson.  In our model 
\begin{equation}
\left.\frac{m_{1^-}}{m_{0^-}}\right|_{m=m_c} = 1.04\,,
\end{equation}
with a splitting of $130\,$MeV, and this splitting drops to just $40\,$MeV at $m_b$; viz., only 5\% of its value in the chiral limit.  In addition to the calculated value, the general pattern of our results argues for the mass of the pseudoscalar partner of the $\Upsilon(1S)$ to lie above $9.4\,$GeV.  Indeed, we expect the mass splitting to be much less than $m_{J/\psi}-m_{\eta_c}$, not more.  (See also; e.g., Ref.\ \cite{fllanes}.)

\begin{figure}[t] 
\vspace*{2em}
 
\centerline{\includegraphics[width=0.45\textwidth]{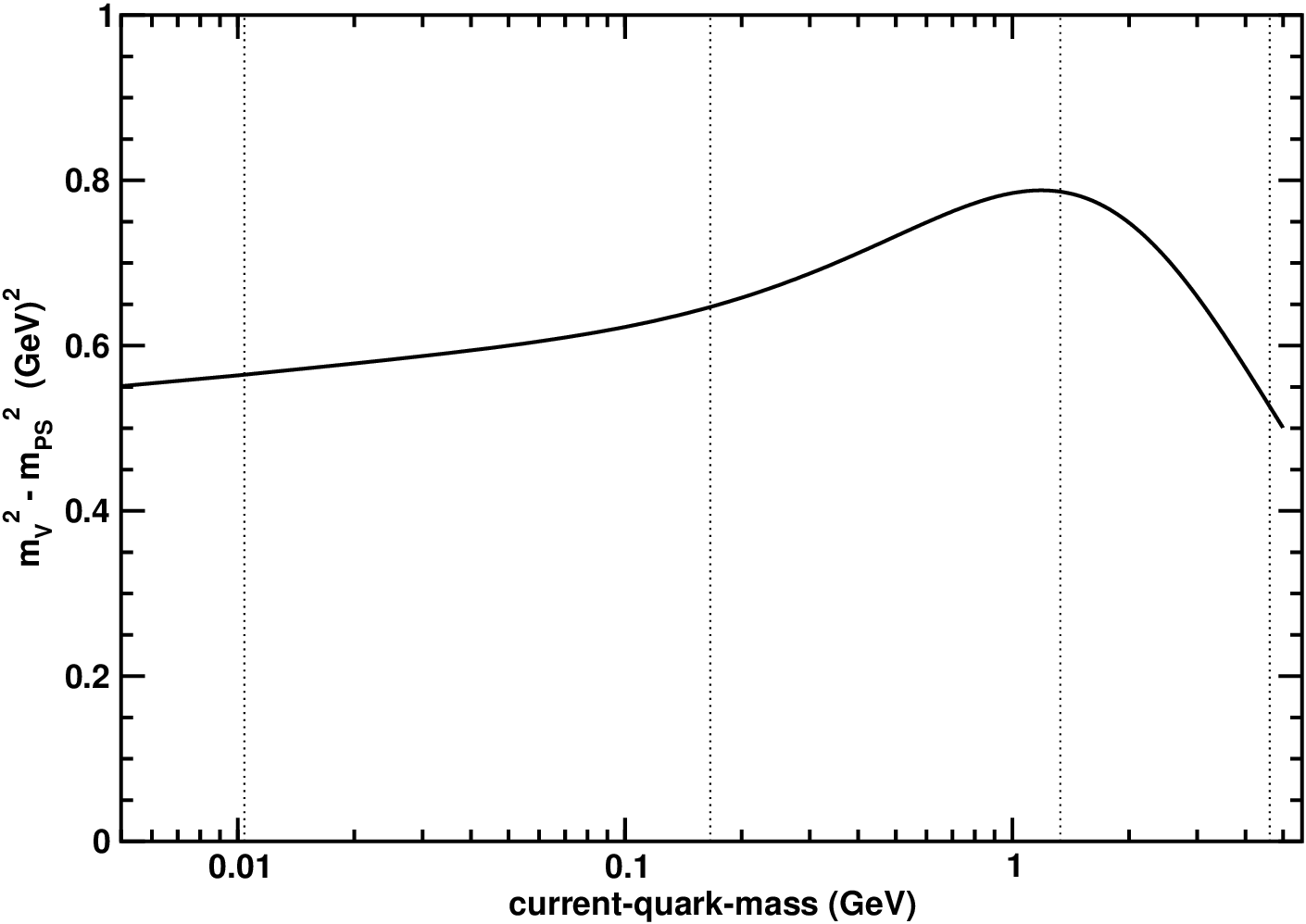}}

\caption{\label{msqdiff} Evolution with current-quark mass of the difference between the squared-masses of vector and pseudoscalar mesons.  These results were obtained with $\bar{\cal C}=0.51$, Eq.\,(\protect\ref{Cbar}), using the completely resummed dressed-quark-gluon vertex in the gap equation and the vertex-consistent Bethe-Salpeter kernel.  The dotted vertical lines mark the current-quark masses in Table \protect\ref{currentquark}.}
\end{figure} 

In Fig.\,\ref{msqdiff} we depict the evolution with current-quark mass of the difference between the squared-masses of vector and pseudoscalar mesons.  On a material domain of current-quark masses this difference is approximately constant: $m_{1^-}^2-m_{0^-}^2 \approx 0.56\,$GeV$^2$, an outcome consistent with experiment that is not reproduced in numerical simulations of quenched lattice-QCD \cite{latticeVPS}.  The difference is maximal in the vicinity of $m_c$, a result which reemphasises that heavy-quark effective theory is not an appropriate tool for the study of $c$-quarks \cite{mishasvy}.

\begin{figure}[t] 
\vspace*{2em}
 
\centerline{\includegraphics[width=0.45\textwidth]{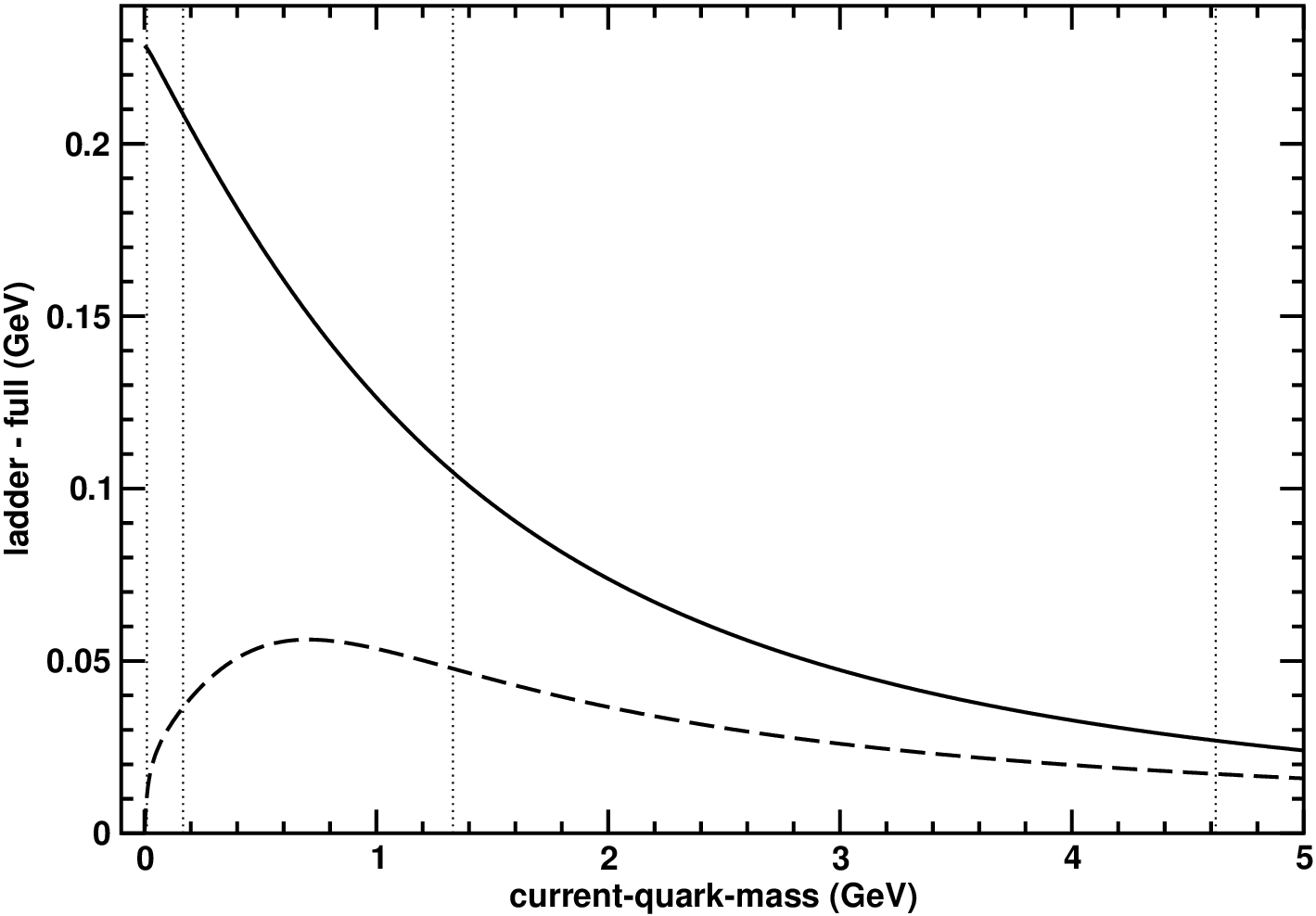}}

\vspace*{3.5em}

\centerline{\includegraphics[width=0.45\textwidth]{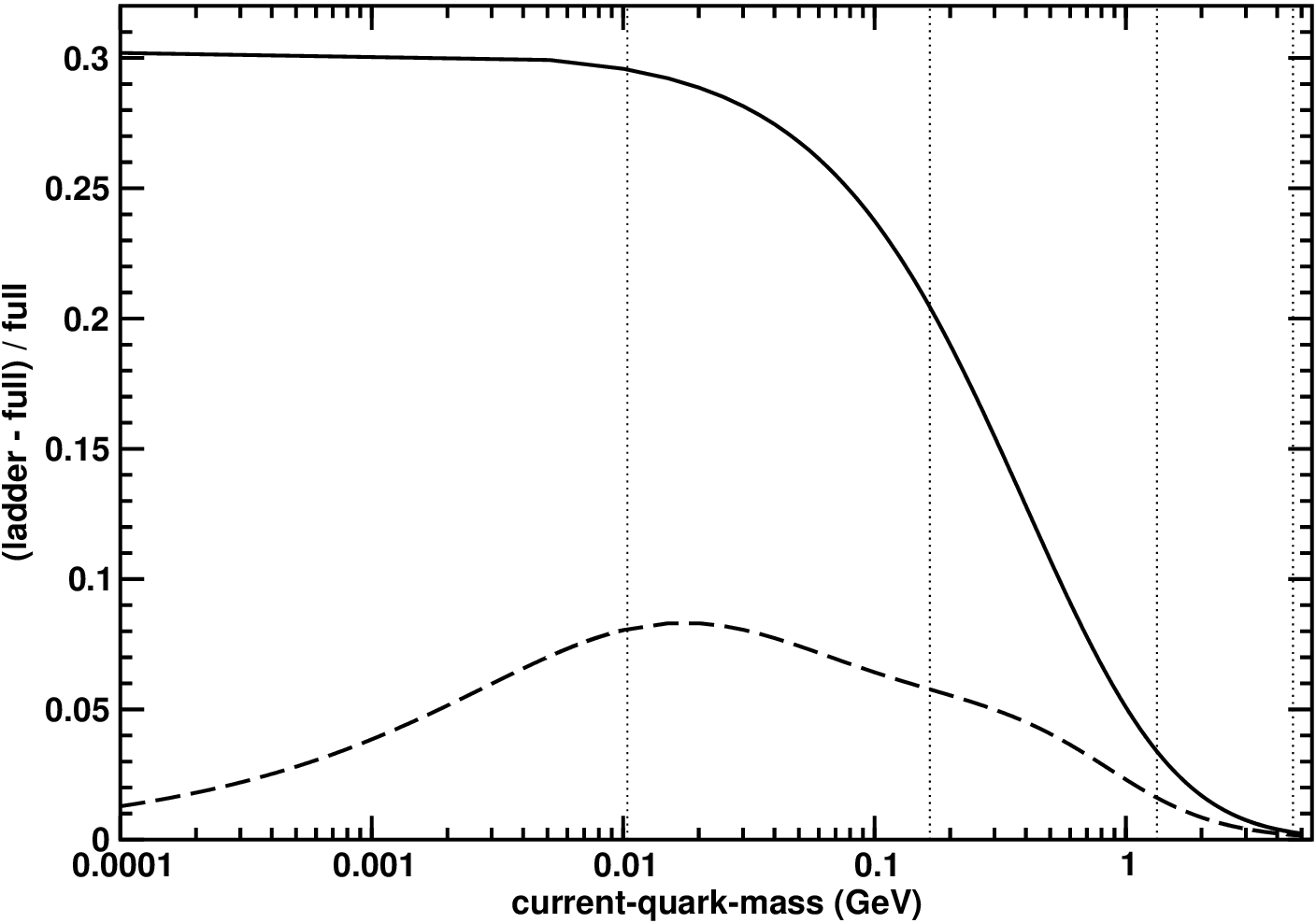}}

\caption{\label{ladderfull} Evolution with current-quark mass of the difference between the meson mass calculated in the rainbow-ladder truncation and the exact value; namely, that obtained using the completely resummed dressed-quark-gluon vertex in the gap equation and the vertex-consistent Bethe-Salpeter kernel.  The \textit{upper panel} depicts the absolute error and the \textit{lower panel}, the relative error.  Solid lines: vector meson trajectories; and dashed-lines; pseudoscalar meson trajectories.   The results were obtained with $\bar{\cal C}=0.51$, Eq.\,(\protect\ref{Cbar}) and the dotted vertical lines mark the current-quark masses in Table \protect\ref{currentquark}.}
\end{figure} 

Figure \ref{ladderfull} is instructive.  It shows that with growing current-quark mass the rainbow-ladder truncation provides an increasingly accurate estimate of the ground state vector meson mass.  At the $s$-quark mass the relative error is 20\% but that has fallen to $< 4$\% at the $c$-quark mass.  Similar statements are true in the valid pseudoscalar channels.  In fact, in this case the agreement between the truncated and exact results is always better; e.g., the absolute difference reaches its peak of $\approx 60\,$MeV at $m\sim 4\,m_s$ whereat the relative error is only 3\%.
This behaviour in the pseudoscalar channel is fundamentally because of Goldstone's theorem, which requires that all legitimate truncations preserve the axial-vector Ward-Takahashi identity and hence give a massless pseudoscalar meson in the chiral limit.  It is practically useful, too, because it indicates that the parameters of a model meant to be employed in a rainbow-ladder truncation study of hadron observables may reliably be fixed by fitting to the values of quantities calculated in the neighbourhood of the chiral limit.  

The general observation suggested by this figure is that with increasing current-quark mass the contributions from nonplanar diagrams and vertex corrections are suppressed in both the gap and Bethe-Salpeter equations.  Naturally, they must still be included in precision spectroscopic calculations.  It will be interesting to reanalyse this evolution in a generalisation of our study to mesons composed of constituents with different current-quark masses, and thereby extend and complement the limited such trajectories in Refs.\,\protect\cite{mariswien,pcterice}.

\subsection{Bethe-Salpeter equation for diquarks}
Colour antitriplet quark-quark correlations (diquarks) have long been a focus of attempts to understand baryon structure \cite{firstdq,lichtenberg}.  It was quickly realised that both Lorentz scalar and axial-vector diquarks, at least, are important for baryon spectroscopy \cite{L2} and, from a consideration of baryon magnetic moments, that diquark correlations are not pointlike \cite{L3}.  An appreciation of their importance has grown and a modern picture of diquark correlations in baryons is realised through their role in a Poincar\'e covariant Faddeev equation \cite{NpiN,regfe,oettel,bentz}.

In rainbow-ladder truncation colour-antitriplet diquarks are true bound states \cite{regdqmass,pieterdq}.  (NB. The other admissible channel; viz., colour sextet, is never bound since even single gluon exchange is repulsive therein.)  In spite of this, the addition of ${\cal L}_2^-(k,p)$, Eq.\,(\ref{L2m}), to the quark-gluon vertex, along with the three terms it alone generates in the colour-antitriplet quark-quark scattering kernel, overwhelms the attraction produced by single gluon exchange and eliminates diquarks from the spectrum \cite{truncscheme,alkoferdq}.  The repulsive effect owing to ${\cal L}_2^-(k,p)$ is consummated when the series it generates is fully resummed \cite{detmold}.  With that vertex, and the consistent colour-antitriplet quark-quark scattering kernel, the characteristic polynomial obtained from the Bethe-Salpeter equation exhibits a pole, which is the antithesis of the zero associated with a bound state.

Nevertheless, herein we have introduced a new element into the consideration of diquark correlations; namely, our model for the colour-octet quark-antiquark scattering kernel exhibits attraction, which is a property not possessed by ${\cal L}_2^-(k,p)$.  The dressed-quarks that appear in the diquark Bethe-Salpeter equation are described by the gap equation we have already elucidated.   However, the manner in which diagrams are combined and resummed in the vertex-consistent colour-antitriplet diquark Bethe-Salpeter kernel is different from that which maintains for colour singlet mesons.  Fortunately, the modifications necessary in the class of \textit{Ans\"atze} containing our model were elucidated in Ref.\ \cite{detmold} and we need only adapt them to our particular case.  

In consequence we arrive at the following Bethe-Salpeter equation:
\begin{eqnarray}
\nonumber
\Gamma_{qq}^{C}(P) &=& -\frac{1}{2}\, \gamma_\mu \, S(Q) \,\Gamma_{qq}^{C}(P) S(-Q) \, \Gamma_\mu^{\cal C}(-Q) \\ 
& & + \frac{3}{4} \, \gamma_\mu \, S(Q) \, {\cal T}_\mu^L(P)\label{BSEqq}
\end{eqnarray}
wherein: $\Gamma_{qq}^{C}(P) = \Gamma_{qq}(P) C^\dagger$, $C= \gamma_2\gamma_4$ is the charge conjugation matrix, with $\Gamma_{qq}(P)$ the Bethe-Salpeter amplitude describing diquark correlations; and $S$ and $\Gamma_\mu^{\cal C}$ are precisely the elements used in the $\pi$- and $\rho$-meson equations.  ${\cal T}_\mu^L(P)$ is new, however.  It is a variant for diquarks of $\Lambda^a_{M\mu}$ in Eq.\,(\ref{Lambdatotal}), as we now show.

The Bethe-Salpeter amplitudes for scalar and axial vector diquarks can be written:
\begin{eqnarray}
\Gamma^{C\, 0^+}_{qq}(P) &= &\gamma_5\,\left[ f_1^{0^+}(P^2) + \gamma\cdot P 
f_2^{0^+}(P^2)\right]\,, \\
\nonumber
\Gamma_{qq}^{C\,1^+}(P) &=& \gamma \cdot\epsilon^\lambda(P)\, f_1^{1^+}(P^2) + 
\sigma_{\mu\nu}\,\epsilon_\mu^\lambda(P)\, \hat P_\nu\, f_2^{1^+}(P^2) \,.\\
\label{gammaavdq} 
\end{eqnarray} 
These amplitudes are identical in structure to those of the $\pi$- and $\rho$-mesons, Eqs.\,(\ref{pimodel}) and (\ref{gammarho}), and that is why we work with $\Gamma_{qq}^{C}$ instead of $\Gamma_{qq}$.  Consequently, the Bethe-Salpeter equations assume the form in Eq.\,(\ref{fHrho})
\begin{equation}
\label{fHqq}
\mbox{\boldmath $f$}_{qq}(P) = {\cal H}_{qq}(P^2) \mbox{\boldmath $f$}_{qq}(P)\,,
\end{equation}
wherein ${\cal H}_{qq}(P^2)$ is a $2\times 2$ matrix: 
\begin{equation}
\label{Hqq}
{\cal H}_{qq}(P^2) = {\cal H}_{qq \Gamma}(P^2) + {\cal H}_{qq \Lambda}(P^2)\,.
\end{equation}
The ${\cal H}_{qq \Gamma}(P^2)$ term is obtained in a straightforward manner from the first line of Eq.\,(\ref{BSEqq}) by following the pattern of Eq.\,(\ref{HGammarho}).  

The second term employs the mapping of $\Lambda^a_{M\mu}$ into the present case, which is
\begin{eqnarray}
\Lambda^{a}_{qq \nu}(P)&=& \sum_{n=0}^{\infty} \Lambda^{a;n}_{qq \nu}(P) 
\end{eqnarray}
with
\begin{eqnarray}
\nonumber \lefteqn{- \frac{1}{6\, {\cal C}}\Lambda^{a;n}_{qq \,\nu}(P)}\\
\nonumber &= & l^b \,\gamma_\rho\, \lambda_\wedge^k \chi_{qq}^C(P)\, 
\Gamma_{\nu,n-1}^{\cal C}(-Q)\,(l^a)^{\rm t} \, S(-Q)\, (l^b)^{\rm t} \gamma_\sigma\\ 
\nonumber & &-\,  l^b \gamma_\rho\, 
S(Q)\,  l^a \Gamma_{\nu ,n-1}^{\cal C}(Q)\, \lambda_\wedge^k \chi_{qq}^C(P) 
\,(l^b)^{\rm t} \gamma_\sigma  \\
 & &- \, l^b \gamma_\rho\, S(Q)\,  \Lambda^{a;n-1}_{qq \,\mu}(P)\, S(-Q) \, (l^b)^{\rm t} \gamma_\sigma \,,
\label{DLambdarecursion} 
\end{eqnarray} 
where the superscript ``t'' denotes matrix transpose,
\begin{equation}
\chi_{qq}^C(P) = S(Q)\, \Gamma_{qq}^{C}(P) \, S(-Q)
\end{equation}
and $\{\lambda_\wedge^1 = l^7,\lambda_\wedge^2 = l^5, \lambda_\wedge^3 = l^2\}$ are three antisymmetric matrices that represent the colour structure of an antitriplet diquark.  For clarity we have already made use of Eq.\,(\ref{mnmodel}).  

In general 
\begin{equation} 
\Lambda^{a;n}_{qq\, \nu}(P) = 
\Lambda^{n}_{1\nu}(P)\,l^a\,\lambda_\wedge^k + 
\Lambda^{n}_{2\nu}(P)\,\lambda_\wedge^k \,(l^a)^{\rm t}\,, 
\end{equation} 
where $\Lambda^{n}_{1\nu}$ and $\Lambda^{n}_{2\nu}$ are Dirac matrices alone.  It follows that
\begin{eqnarray}
\nonumber
 - \sfrac{1}{8 \, {\cal C}} \, {\cal T}^{L;n}_\nu 
\,\lambda_\wedge^k &:=&  - \sfrac{1}{8 \, {\cal C}}
\,l^a\,\Lambda^{a;n}_{D\nu} \\
\nonumber
& = &  \sfrac{1}{12\,{\cal C}} \left[2 
\Lambda_{1\nu}^n - \Lambda_{2\nu}^n \right] \lambda_\wedge^k \\ 
\nonumber
&= & -\frac{5}{12} \, G^{1,n-1}_{qq \nu }(P) - \frac{1}{12} \,  G^{2,n-1}_{qq \nu }(P) \\
\nonumber && - \frac{1}{12}\, L^{1,n-1}_{qq \nu}(P) + \frac{5}{12} \,
  L^{2,n-1}_{qq \nu}(P)\,,\\
\label{TLn}\\
\nonumber
 - \sfrac{1}{8 \, {\cal C}} \, {\cal T}^{R;n}_\nu  \,\lambda_\wedge^k &:=&  
- \sfrac{1}{8 \, {\cal C}} \,\Lambda^{a;n}_{D\nu} \,(l^a)^{\rm t}\\
\nonumber
&= & \sfrac{1}{12\,{\cal C}} \left[- \Lambda_{1\nu}^n + 2\Lambda_{2\nu}^n \right] \lambda_\wedge^k \\ \
\nonumber
&= & \frac{1}{12} \, G^{1,n-1}_{qq \nu }(P) + \frac{5}{12} \,  G^{2,n-1}_{qq \nu }(P) \\
\nonumber && \frac{5}{12}\, L^{1,n-1}_{qq \nu}(P) - \frac{1}{12} \,
  L^{2,n-1}_{qq \nu}(P)\,,\\ \label{TRn} 
\end{eqnarray}
wherein
\begin{eqnarray}
\label{G1n}
G^{1,n}_{qq \nu }(P) &= &\gamma_\rho \,\chi_{qq}^C(P)\, \Gamma_{\nu,n}^{\cal C}(-Q)\, S(-Q) \gamma_\rho\,, \\
\label{G2n}
G^{2,n}_{qq \nu}(P) &= &\gamma_\rho \, S(Q)\, \Gamma_{\nu,n}^{\cal C}(Q)\, \chi_{qq }^C(P)\, \gamma_\rho\,,
\end{eqnarray}
and
\begin{equation}
\label{Lin}
L^{i,n}_{qq \nu}(P)= \gamma_\rho S(Q) \, \Lambda^n_{i \nu}(P) \, S(-Q) \gamma_\rho\,,\; i=1,2\,.
\end{equation}
NB.\ It is apparent from the first lines in each of Eqs.\,(\ref{TLn}), (\ref{TRn}) that there is no mixing between colour antitriplet and colour sextet diquarks.

In this sequence of equations one has arrived at an analogue of Eq.\,(\ref{Leasy}).  Although somewhat more complicated because of the colour-antitriplet nature of the diquarks, the series expresses a recursion relation that may be written compactly as
\begin{eqnarray}
\nonumber
- \frac{1}{\cal C}\left[ 
\begin{array}{c}
\Lambda_{1\nu}^n\\[1ex]
\Lambda_{2\nu}^n
\end{array}
\right] & = &
[
T_G] \,
\left[ 
\begin{array}{c}
G_{qq \nu}^{1,n-1}\\[1ex]
G_{qq \nu}^{2,n-1}
\end{array}
\right] 
+ [T_L]\,
\left[
\begin{array}{c}
L_{qq \nu}^{1,n-1}\\[1ex]
L_{qq \nu}^{2,n-1}
\end{array}
\right] \\
\end{eqnarray} 
with
\begin{eqnarray}
[T_G] = 
\left[
\begin{array}{cc}
1 & -3 \\[1ex]
3 & - 1 
\end{array}
\right]
,\;
[T_L]= 
- \left[
\begin{array}{cc}
3 & 1 \\[1ex]
1 & 3  
\end{array}
\right] ,
\end{eqnarray} 
and which is subject to the initial conditions: $\Lambda_{1,2\, \nu}^0=0$.   NB.\ In this case, owing to modifications associated with the colour algebra, $\Lambda_{0^+ \nu}^a(P)\neq 0$.  Hence both the scalar and axial-vector diquark Bethe-Salpeter equations contain nonplanar contributions generated by the analogue of Eq.\,(\ref{Hrholast}).  

From this point one proceeds for both the scalar and axial-vector diquarks as one did from Eq.\,(\ref{Lambdarho}) for the $\rho$-meson.  The matrices $\Lambda_{1\mu}$ and $\Lambda_{2\mu}$ are expressed in terms of independent vectors $\mbox{\boldmath $\beta$}_1$, $\mbox{\boldmath $\beta$}_2$; viz., 
\begin{equation}
\mbox{\boldmath $\beta$}_{qq i} = M_\nu \Lambda_{i\nu}\,,
\end{equation}
which can be combined into a single column vector $\mbox{\boldmath $\beta$}_{qq}$.  That vector is given by 
\begin{equation}
\label{betaqq}
\mbox{\boldmath $\beta$}_{qq} = \frac{1}{1 - M_\nu\, T_L\, \tilde L_{qq \nu}} \, M_\nu \,T_G\, G_{qq \nu}
\end{equation}
wherein $G_{qq \nu}$ is the matrix constructed from Eqs.\,(\ref{G1n}) and (\ref{G2n}) using the fully resummed quark-gluon vertex; i.e., $\Gamma_{\nu n}^{\cal C} \to \Gamma_{\nu}^{\cal C}$ on the r.h.s.\ in these equations, and $\tilde L_{qq \nu}$ is defined implicitly via Eq.\,(\ref{Lin}); namely, the right-hand-sides are combined into a single matrix and identified as $\tilde L_{qq \nu}\mbox{\boldmath $\beta$}_{qq}$.   With $\mbox{\boldmath $\beta$}_{qq}$ in Eq.\,(\ref{betaqq}) one has in hand a closed form for 
\begin{equation}
{\cal T}_\mu^L(P)= -\frac{2}{3}\,[2\, \Lambda_{1\mu} - \Lambda_{2\mu}]
\end{equation}
in Eq.\,(\ref{BSEqq}) and therefrom ${\cal H}_{qq \Lambda}(P^2)$ in Eq.\,(\ref{Hqq}).  Thus has one arrived finally at the $2\times 2$ matrices  needed to construct analogues of Eq.\,(\ref{rhochar}); viz., two characteristic equations, one for the scalar diquark and another for the axial-vector.  A bound diquark exists if either or both of these equations possess a real zero. 

\begin{figure}[t] 
\vspace*{2em}
 
\centerline{\includegraphics[width=0.45\textwidth]{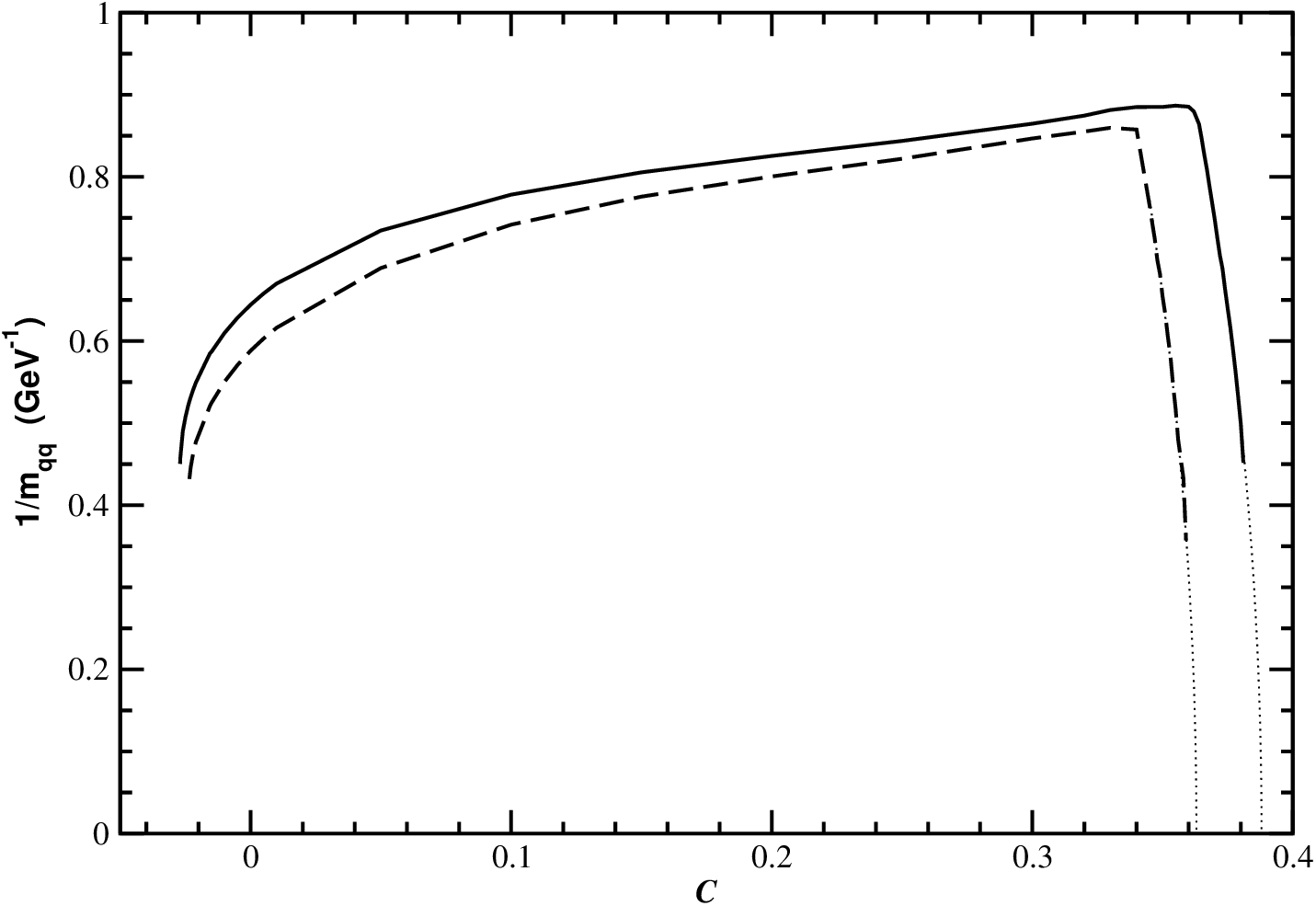}}

\caption{\label{mdqC} Evolution of the reciprocal diquark masses, calculated in the chiral limit, with the parameter ${\cal C}$, a positive value of which expresses attraction in the colour-octet projection of the quark-antiquark scattering kernel: solid line -- scalar diquark, dashed line -- axial-vector diquark.  The dotted lines are explained in connection with Eq.\,(\protect\ref{criticalC}).  The mass-scale ${\cal G} = 0.69\,$GeV.}
\end{figure} 

We have studied the behaviour of these characteristic equations as the parameter ${\cal C}$ in Eq.\,(\ref{L2C}) is varied and our results are summarised in Fig.\,\ref{mdqC}.  A value of ${\cal C}=0$ implements the rainbow-ladder truncation.  In this case, as we promised, both scalar and axial-vector diquarks are bound, with $m_{0^+}=1.55\,$GeV and $m_{1^+}=1.70\,$GeV, in agreement with Ref.\,\cite{truncscheme}.  

When ${\cal C}$ is evolved to negative values, which corresponds to net repulsion in the colour-octet projection of the quark-antiquark scattering kernel, the position of the zero in both characteristic equations moves deeper into the timelike region; i.e., the diquark masses increase.  This continues briefly until, at ${\cal C}^-_{1^+}\approx -0.023 \approx -1/43$, the characteristic equation for the axial-vector diquark no longer has a solution.  For the scalar diquark this happens at ${\cal C}^-_{0^+}\approx -0.027 \approx -1/37$.  It is therefore clear that very little repulsion in the colour-octet quark-antiquark scattering kernel is sufficient to prevent the appearance of diquark bound states.  NB.\ If one considers single-gluon exchange between the quark and antiquark then a value of ${\cal C}= -1/8$ is obtained, Eq.\,(\ref{repulsive}).  This is approximately five-times larger than these critical values, an observation which further elucidates the results in Refs.\ \cite{truncscheme,detmold}.

The evolution of ${\cal C}$ to positive values provides altogether new information and insight.  To begin, the bound diquarks that existed in rainbow-ladder truncation survive and their masses decrease continuously with increasing attraction.  Such behaviour, although not necessarily welcome, might have been anticipated on the basis of continuity, and a decrease in a bound state's mass with increasing attraction between its constituents is not unusual.  This smooth development continues, for the axial-vector diquark until ${\cal C}^+_{1^+}\approx 0.34$ and for the scalar diquark until ${\cal C}^+_{0^+}\approx 0.36$, at which point it changes dramatically.  The masses suddenly begin to increase rapidly and their behaviour thereafter is described by
\begin{equation}
\frac{1}{m_{0^+}} = 5.59 \, \sqrt{0.388 - {\cal C}}\,,\;
\frac{1}{m_{1^+}} = 5.75 \, \sqrt{0.363 - {\cal C}}\,.
\label{criticalC}
\end{equation}
Equations (\ref{criticalC}) are represented by the dotted lines in Fig.\ \ref{mdqC}.  

The value of ${\cal C}$ at which the behaviour of the masses changes qualitatively is correlated with the movement of the cusp evident in Figs.\ \ref{As} -- \ref{a3s} into the domain that affects the position of the zero in the characteristic polynomials.  This means that, with net attraction in the colour-octet quark-antiquark scattering channel, the expulsion of diquarks from the bound state spectrum follows immediately upon the active expression of confinement by the dressed-quark propagator in the bound state equation.  

There are no bound diquarks in the spectrum obtained with the value of ${\cal C}=\bar{\cal C}$ in Eq.\,(\ref{Cbar}) suggested by the lattice data.

\section{Epilogue}
\label{conclusion}
Herein we have explored the character of the dressed-quark-gluon vertex and
its role in the gap and Bethe-Salpeter equations.  Our results are relevant
to the mechanism and realisation of confinement and dynamical chiral
symmetry breaking, and the formation of bound states.

We employed a simple model for the dressed-gluon interaction to build an
\textit{Ansatz} for the quark-gluon vertex whose diagrammatic content is
expressly enumerable.  The model reduces coupled integral equations to
algebraic equations and thereby provides a useful intuitive tool.  We used
this framework to argue that data obtained in lattice simulations of
quenched-QCD indicate the existence of net attraction in the colour-octet
projection of the quark-antiquark scattering kernel.

We observed that the presence of such attraction can materially affect the
uniformity of pointwise convergence to solutions of the gap and vertex
equations.  For example, in the timelike region, the vertex obtained by
summing an odd number of loop corrections is pointwise markedly different
from that obtained by summing an even number of loops.  The two subseries of
vertices so defined follow a different pointwise path to the completely
resummed vertex.  Furthermore, correlated with this effect, we found that
solutions of a gap equation defined using an odd-loop vertex and those of a
gap equation defined with an even-loop vertex follow a different pointwise
path to the solution of the exact gap equation; viz., the gap equation
defined by the fully resummed vertex.  This entails that the solutions of two
gap equations that are defined via vertex truncations or vertex \textit{Ans\"atze} which appear similar at spacelike momenta need not yield qualitatively equivalent results for the dressed-quark propagator.  This is especially true in connection with the manifestation of confinement, for which the behaviour of Schwinger functions at timelike momenta is of paramount importance.

Our study showed that the dependence of the dressed-quark-gluon vertex on the
current-quark mass is weak until that mass becomes commensurate in magnitude
with the theory's intrinsic mass-scale.  For masses of this magnitude and
above, all vertex dressing is suppressed and the dressed vertex is well
approximated by the bare vertex.

It is critical feature of our study that the diagrammatic content of the
model we proposed for the vertex is explicitly enumerable because this enables the systematic construction of quark-antiquark and quark-quark scattering kernels that ensure the preservation of all Ward-Takahashi identities associated with strong interaction observables.  This guaranteed, in particular, that independent of the number of loop corrections incorporated in the dressed-quark-gluon vertex, and thereby in the gap equation, the pion was
automatically realised as a Goldstone mode in the chiral limit.  Such a
result is impossible if one merely guesses a form for the vertex, no matter
how sound the motivation.  As a consequence we could reliably explore the
impact on the meson spectrum of attraction in the colour-octet projection of
the quark-antiquark scattering kernel.  In accordance with intuition, the
mass of a meson decreases with increasing attraction between the
constituents.

We found that the fidelity of an approximate solution for a meson's mass
increases with the number of loops retained in building the vertex.  However,
a given consistent truncation need not yield a solution.  This fact is tied
to a difference between the convergence paths followed by the odd-loop
vertex series and the even-loop series.  In addition, we observed that with
increasing current-quark mass the rainbow-ladder truncation provides an ever
more reliable estimate of the exact vector meson mass; i.e., the mass
obtained using the completely resummed vertex and the completely consistent
Bethe-Salpeter equation.  For pseudoscalar mesons, this is even more true
because the rainbow-ladder and exact results are forced by the Ward-Takahashi
identity to agree in the chiral limit.  This is practically useful because it
means that the parameters of a model meant to be employed in rainbow-ladder
truncation may reliably be fixed by fitting to the values of pseudoscalar
meson quantities calculated in the neighbourhood of the chiral limit.
Moreover, both in rainbow-ladder truncation and with the complete vertex and
kernel, the splitting between pseudoscalar and vector meson masses vanishes
as the current-quark mass increases.  In our complete model calculation this splitting is $130\,$MeV at the $c$-quark mass and only $40\,$MeV at the $b$-quark mass, a pattern which suggests that the pseudoscalar partner of the $\Upsilon(1S)$ cannot have a mass as low as that currently ascribed to the $\eta_b(1S)$.

Our framework permitted a treatment of diquark correlations on precisely the
same footing as mesons.  In this case we exposed particularly interesting
consequences of attraction in the colour-octet projection of the
quark-antiquark scattering kernel.  Colour-antitriplet diquark correlations
form bound states in rainbow-ladder truncation.  We demonstrated, however, that introducing a very small amount of repulsion into the kernel eliminates these states from the strong interaction spectrum.  If, on the other hand, one
introduces a small amount of colour-octet attraction, these diquarks persist
as bound states.  However, as one increases the amount of attraction toward
that indicated by lattice simulations a dramatic change occurs.  The diquark masses suddenly begin to increase rapidly when the amount of attraction is sufficient to cause the confining nonanalyticity of the dressed-quark propagator to enter into the domain which affects binding in
the colour-antitriplet channel.  The growth continues and, with little further increase in attraction, diquarks vanish from the spectrum.  (NB.\ There are no bound diquarks in the spectrum obtained with the amount of attraction suggested by lattice data.)  This occurs because the colour algebra in the antitriplet channel acts to screen attraction between the quarks in a diquark correlation so that, with sufficient strength in the gap equation, dressed-quark confinement alone becomes the dominant feature and diquarks are thereby expelled from the bound state spectrum.

It would be natural and useful to extend this work to systems composed of
quarks with unequal current-quark masses.

\begin{acknowledgments} 
We are pleased to acknowledge valuable interactions with R.\ Alkofer, A.\ {K\i z\i lers\"u} and F.\,J.\ Llanes Estrada.
This work was supported by: the Austrian Research Foundation \textit{FWF, 
Erwin-Schr\"odinger-Stipendium} no.\ J2233-N08; the Department of Energy, 
Nuclear Physics Division, contract no.\ W-31-109-ENG-38; the \textit{A.v.\ 
Humboldt-Stiftung} via a \textit{F.W.\ Bessel Forschungspreis}; and the National Science
Foundation under contract nos.\ PHY-0071361 and INT-0129236.
\vspace*{\fill}
\end{acknowledgments} 

\end{document}